\documentclass[3p]{elsarticle}

\usepackage[utf8]{inputenc}
\usepackage{amsmath, amssymb}
\usepackage{hyperref}
\usepackage{enumitem}
\usepackage{graphicx} 
\usepackage{float}     
\usepackage{booktabs}
\usepackage{multirow}
\usepackage{natbib}
\usepackage{orcidlink}
\usepackage{bm}

\hypersetup{colorlinks=true, linkcolor=blue, urlcolor=blue, citecolor=blue}

\begin{document}

\begin{frontmatter}

\title{Interaction of Sustained Vortex Shedding with a Phononic Subsurface: Effect of Positioning and Resonant Characteristics}

\author[1]{Sangwon Park\orcidlink{0009-0009-0129-5414}}
\author[1]{Vinod Ramakrishnan\orcidlink{0000-0002-5588-876X}}
\author[2]{Arturo Machado Burgos\orcidlink{0000-0001-8327-7790}}
\author[2]{Andres Goza\orcidlink{0000-0002-9372-7713}}
\author[1]{Kathryn H. Matlack\orcidlink{0000-0001-7387-2414}}

\affiliation[1]{organization={The Grainger College of Engineering, Department of Mechanical Science and Engineering},
            addressline={University of Illinois Urbana-Champaign},
            city={Urbana},
            postcode={61801},
            state={Illinois},
            country={USA}}

\affiliation[2]{organization={The Grainger College of Engineering, Department of Aerospace Engineering},
            addressline={University of Illinois Urbana-Champaign},
            city={Urbana},
            postcode={61801},
            state={Illinois},
            country={USA}}

\begin{abstract}
Phononic materials (PMs) have emerged as a promising candidate to passively and beneficially interact with unsteady flows. However, their fluid--structure interaction (FSI) with separated flows exhibiting sustained periodic vortex shedding remains completely unexplored. In this study, we use a high-fidelity immersed boundary method to systematically investigate the strongly coupled, nonlinear FSI dynamics of vortex shedding from a fully separated flow over a flat plate that exhibits sustained periodic vortex shedding, with an embedded PM subsurface (PSub) modeled as a grounded diatomic chain driving a compliant section on the flat plate. We probe the dependence of the coupled system dynamics on two parameters: (1) the PSub chordwise position, which sets its spatial relationship to the surrounding flow structure, and (2) the rate of growth of the resonant displacement amplitude of a PSub, a behavioral parameter identified as a key driving parameter in PSub interaction with latent vortex shedding in prior work. Simulation results show that these two parameters drive clear and systematic changes to the limit-cycle vortex shedding. The PSub chordwise position determines the limit cycle PM--FSI dynamics, which can be classified into four distinct regimes as a function of position, ranging from substantial amplification to strong attenuation of the lift and drag oscillation amplitudes in the limit cycle. Results reveal a configuration that reduces lift oscillation amplitude by over 30\% and drag oscillation amplitude by over 50\% with less than 0.5\% change in mean lift and drag. Displacement amplitude envelope primarily governs the transient responsiveness toward the limit cycle, where its influence on the limit cycle also depends on the chordwise position. We analyze the characteristic frequencies, phase relationship between the PSub's displacement and the lift, and unsteady and mean flow fields of the vortex-shedding process to explain the mechanisms by which the PM modifies the flow.
\end{abstract}


\end{frontmatter}

\section{Introduction}

The interaction between phononic materials (PMs) and fluid flows represents an emerging area of fluid--structure interaction (FSI) research, with promising implications for passive, adaptive flow control. PMs are periodic structures whose spatially periodic material, geometric, and/or boundary conditions cause wave propagation features that are not possible in conventional structures. The periodicity produces frequency-dependent wave propagation and vibration characteristics, including band gaps and truncation resonances (TRs) \citep{hussein2015flow,ParkJASA2024}. While PMs have been extensively studied in the context of elastic wave control and vibration isolation, their coupling to surrounding fluid flows and the capability to passively alter those flows is still an emerging area of research \cite{AvalloneNC2026}. A fundamental understanding of how PMs interact with flows, how their structural parameters affect the coupled FSI dynamics, and how these parameters can be systematically tuned to achieve desired flow outcomes is needed in order to incorporate PMs into engineering systems for passive flow control.

\citet{hussein2015flow} pioneered the use of a phononic subsurface (PSub), integrated into the bottom wall of a channel flow, to interact with Tollmien-Schlichting (TS) wave instabilities and numerically demonstrated the PSub response can destructively interfere with and locally attenuate the TS wave at the interface. Here, the PSub parameters were tuned to exhibit a ``truncation resonance'', which is a structural resonance in the band gap of the PMs~\citep{BastawrousJASA2022,HasanJEL2024,ParkJASA2024}. This resonance was tuned to align close to the frequency of the TS wave and respond out of phase, resulting in local phase cancellation of the TS wave instability of the flow.
This effort motivated a number of studies in PM-FSI (see, e.g.,~\citet{AvalloneNC2026} for a review). Several contributions followed to better understand the local interaction between TS waves and the out-of-phase response of the truncation resonance (\citet{BarnesAIAA2021,KianfarNJP2023,KianfarJAP2023, MichelisPoF2023}), though translating this local interaction to a global delay in the transition remained a challenge. \citet{WilleyJFS2023} overcame limitations of single interaction surfaces and showed that a multi-input multi-output FSI configuration can interact with a seeded TS wave in the flow, to passively achieve the required positive force-displacement phasing that leads to successful transition delay.
Other interaction approaches have shown that a PSub acoustic diode can extract energy from a laminar flow of a TS wave to mitigate these instabilities and delay transition \citet{SchmidtJAP2025}, and that band gaps can attenuate traveling wave flutter instabilities that instigate within compliant wave surfaces that beneficially interact with TS waves \citet{FabbianeJFM2025}. 

Beyond TS wave interactions with PSubs, other contributions have extended PM-based interaction to other flow configurations through distinct mechanisms. \citet{NavarroMatter2025} mitigated the unsteadiness of a hypersonic shock-wave/boundary layer interaction using a bilayer PSub whose band gap spans the frequency range where the unsteadiness is most severe. In a related vein, \citet{LinArXiV2026} leveraged a ``defect resonance''~\citep{RamakrishnanJSV2025}, another type of structural resonance in the phononic band gap of the PSub, to interact with spatio-temporally broadband turbulent flow structures within a wall-bounded turbulent boundary layer to achieve turbulent drag reduction.

PSubs have also been used to interact with separated flows exhibiting vortex shedding. \citet{KeoghArXiV2025} experimentally modified the shedding process induced by an upstream wire using a thin locally resonant metamaterial subsurface that has polarization-dependent band gaps, where an amplified out-of-plane surface response at the target shedding frequency interferes destructively with vortex development. Most relevant to the present work, recent foundational work by the authors showed that the phononic material parameters could be recast into FSI-relevant ``behavioral'' parameters, enabling a systematic approach to studying the fully coupled FSI with a band gap resonance of a phononic material~\citep{RamakrishnanJFS2026}.

There, the authors proposed four FSI-relevant PM behavioral parameters, the effective static stiffness $k_\mathrm{eff}$, the truncation resonance frequency $f_\mathrm{TR}$, the displacement amplitude envelope $\lambda$, and the unit cell mass $m_\mathrm{UC}$, as a parameterization of the PM distinct from its raw structural parameters, chosen to enable a systematic study of how the PSub modulates the flow. The authors investigated PM-FSI using simulations of an aerodynamic flow ($Re=400$) over a PSub-embedded flat plate inclined at a $12^\circ$ angle of attack.
For these parameters, the baseline flow---fully rigid with no PSub---exhibits \emph{latent} vortex shedding, where the rigid baseline flow is fully steady but near the onset of vortex shedding~\citep{RamakrishnanJFS2026}. That is, the flow has small trailing edge separation but is steady; but a small increase in angle of attack or Reynolds number triggers indefinite periodic vortex shedding. Introducing the PSub--triggered unsteadiness, and the results established the impact of the behavioral parameters on the single-/broad-band frequency content, wavelength and vorticity of the FSI-induced vortex shedding, and on the mean and dynamic lift force. Specifically, the displacement amplitude envelope of the phononic subsurface, a parameter previously overlooked in PM--FSI studies, emerged alongside $f_\mathrm{TR}$ as one of the key driving parameters in this latent vortex shedding FSI. 

We build on this prior work by investigating not how  PSub interacts with a flow near the onset of unsteady separation, but at a slightly higher $15^\circ$ angle of attack where the baseline flow (i.e., over a rigid plate) is fully separated with sustained periodic vortex shedding. In this setting, we quantify how an important subset of the PM behavioral parameters proposed in~\citep{RamakrishnanJFS2026} affect the underlying unsteady separation process. Because of the location-specific nature of vortex formation along the plate, we probe an additional key parameter: the position of phononic subsurface on the angled flat plate. Specifically, we determine the dependence of the nonlinear FSI on two parameters: $X_c$, which defines the PSub's position along the flat plate, and $\lambda$, which governs the PSub's displacement response at resonance, characterizing how these two parameters govern the limit cycle amplification or attenuation of the vortex shedding and the mechanisms underlying this behavior. The remaining behavioral parameters ($k_{\mathrm{eff}}$, $m_{\mathrm{UC}}$, $f_{\mathrm{TR}}$) are held fixed throughout. $k_{\mathrm{eff}}$ is fixed to keep the mean static displacement of the PSub constant across cases to isolate the effect of the dynamic response on the separated flow. $m_{\mathrm{UC}}$ is fixed as it was found to play a comparatively less consequential role except at low $\lambda$ in the prior study. We fix $f_{\mathrm{TR}}$ aligned with the vortex shedding frequency of the rigid baseline flow as the prior study showed aligning $f_{\mathrm{TR}}$ with the underlying vortex shedding gave the largest coupled response.

The remainder of the paper is organized as follows. Section \ref{sec:problem} describes the problem configuration, computational framework, and PSub model. Section \ref{sec:position} presents the dependence of limit cycle PM--FSI dynamics on PSub chordwise position, showing that $X_c$ determines the limit cycle PM--FSI behavior and reveals four distinct regimes along the chord ranging from significant amplification to strong attenuation of unsteady loads from vortex shedding. Section \ref{sec:lambda} presents dependence of limit cycle PM--FSI dynamics on PSub $\lambda$ at four representative positions, demonstrating that $\lambda$ governs the structural displacement response and the transient evolution of the FSI, with an additional position-dependent influence on the limit cycle. Section \ref{sec:conclusions} summarizes the findings and discusses their implications for passive flow control and PM-based FSI applications.

\section{Problem Formulation}
\label{sec:problem}

\subsection{Flow Configuration and Geometry}
\label{subsec:flow_config}

\begin{figure}[H]
    \centering
    \includegraphics[width=0.9\textwidth]{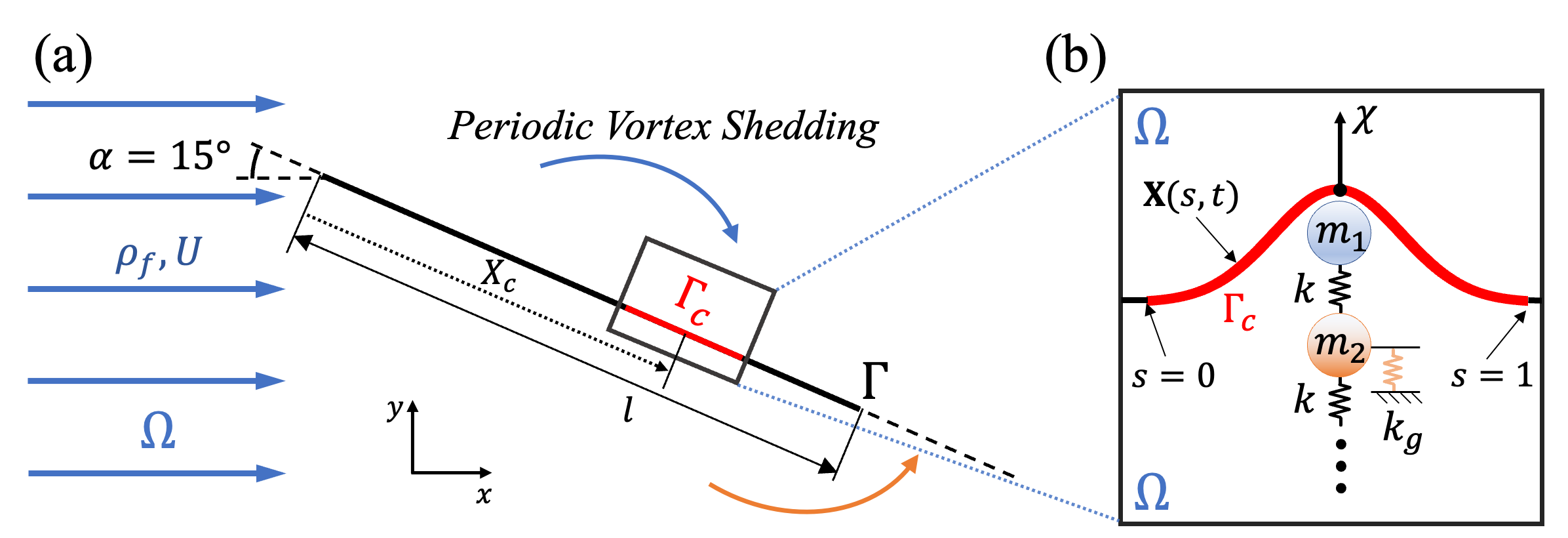}
    \caption{Schematics of the configuration: (a) two-dimensional flow over a flat plate of chord length $l$ at angle of attack $\alpha = 15^\circ$ in a uniform freestream of velocity $U$, with the compliant section $\Gamma_c$ embedded at chordwise position $X_c$ within the otherwise rigid plate surface $\Gamma$; (b) detail of the PSub: a grounded diatomic chain with alternating masses $m_1, m_2$ and springs $k, k_g$, where the displacement $\chi$ of the interface mass drives a Gaussian deformation of the compliant section.}
    \label{fig:problem_config}
\end{figure}

We consider two-dimensional incompressible flow over a flat plate whose conditions induce sustained vortex-shedding instabilities. Specifically, we consider an infinitely thin flat plate of chord length $l$ inclined at an angle of attack $\alpha = 15^\circ$ in a uniform freestream of velocity $U$ and density $\rho_f$, as illustrated in Fig.~\ref{fig:problem_config}(a). The chord-based Reynolds number is set to $Re = U l / \nu = 400$, where $\nu$ is the kinematic viscosity of the fluid. At this angle of attack and Reynolds number, the rigid baseline flow is fully separated from the leading edge and exhibits sustained periodic vortex shedding. Prior work by the authors studied a flat plate at $\alpha = 12^\circ$, where the rigid baseline flow is fully steady but near the onset of vortex shedding~\citep{RamakrishnanJFS2026}, which we refer to as \emph{latent} vortex shedding. This prior work aimed to understand how the PSub-FSI triggers this vortex shedding and alters the resulting FSI behaviors. The purpose of the present work is to quantify how a PSub modulates the vortex-shedding process in a fully separated flow even with a rigid flat plate.


A finite compliant section $\Gamma_c$ is embedded within the otherwise rigid plate surface $\Gamma$, occupying 10\% of the chord length centered at the chordwise position $X_c$. The chordwise position $X_c$ is defined as the streamwise location of the center of $\Gamma_c$ normalized by the chord length and is swept in this work. The remainder of the plate surface is treated as rigid. The compliant section deforms in the wall-normal direction with respect to the plate angle, following a prescribed Gaussian profile, where the deformation amplitude of the midpoint of the compliant section is dictated by the displacement $\chi(t)$ of the interface mass of the PSub (Fig.~\ref{fig:problem_config}(b)).

\subsection{Computational Framework}
\label{subsec:numerical}

We use the strongly coupled immersed-boundary projection framework of \citet{goza2017strongly} to solve the coupled dynamics of the incompressible flow, immersed flat plate, and phononic subsurface. This approach was also used in~\citet{RamakrishnanJFS2026}, where the framework was extended to account for the movement of the compliant section induced by the PSub motion. We adopt the same modeling framework here, including the same PSub model and compliant-section coupling; the following therefore only retains the elements essential to the present configuration, and we refer the reader to~\citet{RamakrishnanJFS2026} for full derivation details. Therefore, we summarize only the elements needed for the present configuration. In this formulation, the aerodynamic loading, PSub response, compliant-section motion, marker redistribution, and no-slip constraint are solved in a coupled manner at each time step.

All quantities are nondimensionalized using the chord length $l$, freestream velocity $U$, fluid density $\rho_f$, and convective time scale $l/U$.  We denote the dimensionless fluid velocity and pressure by $\mathbf{u}$ and $p$, respectively. The immersed-boundary surface stress that enforces the presence of the plate is denoted by $\mathbf{f}$, and the instantaneous plate-surface position is denoted by $\mathbf{X}$. The dimensionless governing equations are
\begin{gather}
\frac{\partial \mathbf{u}}{\partial t}
+\mathbf{u}\cdot\nabla\mathbf{u}
=
-\nabla p
+\frac{1}{Re}\nabla^2\mathbf{u}
+\int_{\Gamma}\mathbf{f}(\tilde{s},t)\,
\delta\left(\mathbf{X}(s(\tilde{s},t),t)-\mathbf{x}\right)\,d\tilde{s},
\label{eqn:NS}
\\
\nabla\cdot\mathbf{u}=0,
\label{eqn:continuity}
\\
\mathbf{M}\frac{d^2\boldsymbol{\chi}}{dt^2}
+\mathbf{K}\boldsymbol{\chi}
=
\mathbf{f}_{\mathrm{PSub}}(t),
\label{eqn:psub_dynamics}
\\
\int_{\Omega}\mathbf{u}(\mathbf{x},t)\,
\delta\left(\mathbf{x}-\mathbf{X}(s(\tilde{s},t),t)\right)\,d\mathbf{x}
=
\frac{d\mathbf{X}(s(\tilde{s},t),t)}{dt},
\qquad \forall \; \mathbf{X}\in\Gamma .
\label{eqn:no_slip}
\end{gather}

Equation~\eqref{eqn:NS} is the incompressible Navier--Stokes momentum equation with the immersed-boundary forcing term. Equation~\eqref{eqn:continuity} enforces mass conservation. Equation~\eqref{eqn:psub_dynamics} gives the equations of motion for the PSub where the vector $\bm{\chi}=[\bm{\chi}_1;\bm{\chi}_2;\cdots;\bm{\chi}_N]$ is the collective displacement of all unit cell masses ($N$: number of unit cells) and $\mathbf{M}$, $\mathbf{K}$ are the mass and stiffness matrices (\ref{app:mk_matrix}). Eq.~\eqref{eqn:no_slip} enforces the no-slip constraint on the immersed plate surface. Here $\Gamma$ is the full plate surface, $s$ denotes the undeformed arc-length coordinate along the compliant section, and $\tilde{s}$ denotes the corresponding  deformed arc-length coordinate. Further details on the immersed-boundary formulation and coordinate transformation between $s$ and $\tilde{s}$ are given elsewhere~\citet{RamakrishnanJFS2026, goza2017strongly}.

The PSub is coupled to the flow through the finite compliant section $\Gamma_c\subset\Gamma$. We parametrize $\Gamma_c$ using a local nondimensional coordinate $s\in[0,1]$, where $s=0$ and $s=1$ correspond to the upstream and downstream endpoints of the compliant section. The deformation of $\Gamma_c$ is fully determined by the displacement of the interface mass of the PSub, $\chi(t)$. Following the localized surface deformation model used in \citet{RamakrishnanJFS2026}, the position of the compliant section is written as

\begin{equation}
    \mathbf{X}(s,t) = \mathbf{X}_{0}(s) + \chi(t)g(s)\hat{\mathbf{n}}(s), \qquad s\in\Gamma_c .
    \label{eqn:gaussian_deformation}
\end{equation}

Here, $\mathbf{X}_{0}(s)$ is the undeformed plate surface position, $\hat{\mathbf{n}}(s)$ is the local unit normal to the plate, and $g(s)$ is a smooth unit-peak Gaussian profile localized over $\Gamma_c$. In the present simulations we use

\begin{equation}
    g(s) = \exp\left[ - \frac{\left(3(2s-1)\right)^2}{2}\right],
    \label{eqn:gaussian_profile}
\end{equation}
so that the maximum normal displacement occurs at the center of the compliant section. Outside $\Gamma_c$, the plate position remains fixed. 

The aerodynamic load applied to the PSub is obtained by integrating the normal component of the immersed-boundary surface stress over the compliant section, 
\begin{equation}
    F_c(t) = \int_{\Gamma_c} \mathbf{f}(\tilde{s},t)\cdot\hat{\mathbf{n}}  (\tilde{s},t)\,d\tilde{s},
    \label{eqn:cs_force}
\end{equation}
where this load enters the PSub equations through the forcing vector, such that only the interface mass is directly forced by the surrounding flow. The resulting interface displacement $\chi(t)$ updates the Gaussian deformation in Eq.~\eqref{eqn:gaussian_deformation}, thereby closing the two-way coupling between the fluid loading, PSub dynamics, and immersed-boundary motion.

Equations~\eqref{eqn:NS}--\eqref{eqn:no_slip} are then spatially discretized using standard second-order finite-difference operators and recast in a vorticity--streamfunction formulation. Boundary conditions on the outer flow domain are handled using a multi-domain approach~\citep{colonius2008fast}. The nonlinear convective term is advanced with an Adams--Bashforth scheme, while the diffusive term is treated implicitly. The PSub dynamics are advanced using a second-order implicit Newmark scheme. The immersed-boundary forcing, no-slip constraint, and PSub degrees of freedom are treated implicitly in time, producing a strongly coupled nonlinear algebraic system at each time step.

The FSI simulations are conducted on a Cartesian grid with uniform spacing $\Delta x = 0.005$ in the region surrounding the plate. Time integration uses a fixed time step of $\Delta t = 0.001$, corresponding to a Courant-Friedrichs-Lewy (CFL) number of approximately 0.2. The plate surface is discretized with a Lagrangian body-point spacing of $\Delta s/\Delta x = 2$ relative to the Eulerian grid along the rigid portion and $\Delta s/\Delta x = 1$ along the compliant section. These parameters correspond to Grid 2 in the grid convergence study presented in ~\ref{app:grid_convergence} (Table~\ref{tab:grid_convergence}), which demonstrates that the mean lift coefficient and vortex shedding frequency obtained at this resolution are converged to within 1\% of the finest grid tested.

\subsection{PSub Model and Behavioral Parameters}
\label{subsec:pm_model}

The phononic subsurface is modeled as a grounded diatomic chain consisting of alternating lumped masses $m_1$ and $m_2$ connected by inter-mass springs of stiffness $k$. Each mass is additionally connected to a rigid ground through grounding springs. The top mass of the chain serves as the interface mass; its vertical displacement $\chi(t)$ governs the Gaussian deformation of the compliant section $\Gamma_c$. The grounded diatomic chain with five unit cells is adopted as a representative model for this study, following the framework previously introduced in \citep{RamakrishnanJFS2026}. This model supports a truncation resonance (TR) as its fundamental mode, with its modal characteristics systematically tuned through the mass and stiffness parameters. Enforcing the TR as the fundamental mode of the structure is an important aspect for FSI in the context of lifting bodies, where the total force that includes lift primarily excites the fundamental mode of the structure (\citep{RamakrishnanJFS2026}).



The PSub is parametrized using the FSI-relevant behavioral parameter framework introduced in \citep{RamakrishnanJFS2026}, which characterizes the PSub through four behavioral parameters rather than the individual masses and stiffnesses of the diatomic chain. The effective static stiffness, $k_\mathrm{eff}$, sets how much the PSub deflects under the mean aerodynamic load. The truncation resonance frequency, $f_\mathrm{TR}$, is the frequency at which the PSub exhibits a resonance with a localized mode shape at the interface, and can be placed above, at, or below the vortex shedding frequency, determining how strongly the PM--FSI synchronize. The displacement amplitude envelope, $\lambda$, quantifies how quickly the PSub's oscillation amplitude grows when excited at the resonant frequency. The unit cell mass, $m_\mathrm{UC}$, is the total mass of a single repeating unit of the PSub, which sets the frequencies of its other structural resonances relative to $f_\mathrm{TR}$. 
These parameters highlight a distinct approach from what is used in designing PSubs for control of laminar-to-turbulent transition (\citet{hussein2015flow,BarnesAIAA2021,KianfarNJP2023,KianfarJAP2023, MichelisPoF2023,WilleyJFS2023}). In that setting, the inlet condition of the flow essentially sets a global flow response at the frequency of the least damped eigenvector of the rigid body configuration. A key control aim, then, is to design the PSub to respond at that frequency, but with a phase offset that leads to attenuation of the instability mode. In the current nonlinear aerodynamic flow, \citet{RamakrishnanJFS2026} demonstrated that the global coupled system can exhibit not just distinct phases but new frequencies. This outcome prompted our avoidance of phase as a key variable for more general coupled fluid-structure interaction systems. Instead, the four behavioral parameters adopted here, $k_{\mathrm{eff}}$, $f_{\mathrm{TR}}$, $\lambda$, and $m_{\mathrm{UC}}$, extend the representative mass and stiffness parameters which has been commonly used in the FSI community ~\citep{williamson2004vortex,sarpkaya2004critical,RamakrishnanJFS2026}, adapted to capture the distinct structural resonances and localized response available to a PSub.

In addition to these four parameters, the PSub chordwise position $X_c$ is introduced as a geometric parameter, setting the PSub's spatial relationship to the surrounding flow. In this study, $\lambda$ and $X_c$ are the two independent parameters, with $k_{\mathrm{eff}}$, $f_{\mathrm{TR}}$, and $m_{\mathrm{UC}}$ held fixed as explained earlier. Unlike the latent vortex shedding of the prior study, where the shedding is triggered by the PSub, the sustained shedding considered here has prominent characteristic frequency without the PSub existence. We align $f_{\mathrm{TR}}$ with this characteristic shedding frequency as a starting point to investigate the PSub's interaction with the sustained vortex shedding, leaving the investigation of detuned configurations, where $f_{\mathrm{TR}}$ does not match the shedding frequency, to future work. $\lambda$ is chosen as the sweep parameter, as it was observed in the prior study to dictate the strength of the resulting FSI coupling when the frequencies were aligned. 

Throughout 54 simulations, we systematically sweep the chordwise position of the PSub and its displacement amplitude envelope. We use a two-stage sweep rather than a full joint sweep over $X_c$ and $\lambda$. In the first stage, we sweep the chordwise position $X_c$ from $0.06$ to $0.86$ in increments of $0.1$ at a fixed, high value of $\lambda$, giving nine PSub positions across the chord. We fix $\lambda$ high in this stage so the PSub couples strongly with the forcing from the flow, making the resulting dependency on $X_c$ more pronounced. This sweep reveals four representative positions that span the observed range of behavior: $X_c = 0.16$ (upstream attenuation), $X_c = 0.36$ (upstream amplification), $X_c = 0.46$ (downstream amplification), and $X_c = 0.66$ (downstream attenuation). The displacement amplitude envelope $\lambda$ is then swept over 6 values at each of these four positions, to test whether this position dependence holds across $\lambda$ ranging from $\lambda/\lambda_{\mathrm{ref}}=2/7$ to $\lambda/\lambda_{\mathrm{ref}}=1$, for a reference value $\lambda_{\mathrm{ref}} = 0.3$. The corresponding mass and stiffness values of the diatomic chain with 5 unit cells at each $\lambda$ are listed in Table~\ref{tab:mk_matrix}; these values are chosen such that $k_{\mathrm{eff}}$, $f_{\mathrm{TR}}$, and $m_{\mathrm{UC}}$ remain constant across the $\lambda$ sweep, isolating the effect of $\lambda$ on the FSI behavior.

\begin{table}[H]
    \centering
    \caption{Behavioral and structural parameters of the PSub with five unit cells. ($k_{\mathrm{eff}}=5.867$,  $f_{\mathrm{TR}}=0.5989$, and $m_{\mathrm{UC}} = 2$)}
    \label{tab:mk_matrix}
    \begin{tabular}{l l c c c c}
        \toprule
        \multicolumn{6}{l}{\textbf{$f_{\mathrm{TR}} = f_{\mathrm{VS}} = 0.5989$}} \\
        \midrule
        \multicolumn{2}{l}{\textbf{Disp. Amp. Env., $\lambda$}} & \textbf{$m_1$} & \textbf{$m_2$} & \textbf{$k$} & \textbf{$k_g$} \\
        \midrule
        $\lambda_{\mathrm{ref}}$    & \textbf{0.300} & 0.382 & 1.618 & 6.530 & 54.264 \\
        $6\lambda_{\mathrm{ref}}/7$ & \textbf{0.257} & 0.343 & 1.656 & 6.770 & 40.620 \\
        $5\lambda_{\mathrm{ref}}/7$ & \textbf{0.214} & 0.309 & 1.690 & 6.910 & 35.863 \\
        $4\lambda_{\mathrm{ref}}/7$ & \textbf{0.171} & 0.273 & 1.727 & 7.000 & 33.110 \\
        $3\lambda_{\mathrm{ref}}/7$ & \textbf{0.129} & 0.232 & 1.768 & 7.070 & 31.179 \\
        $2\lambda_{\mathrm{ref}}/7$ & \textbf{0.086} & 0.183 & 1.818 & 7.130 & 29.803 \\
        \bottomrule
    \end{tabular}
\end{table}

\section{Dependence of Lift and Drag Amplification or Attenuation on PSub Position}
\label{sec:position}
We first characterize how the PSub chordwise position $X_c$ alters the post-bifurcation vortex shedding response at $\lambda=0.3$ used in the first sweep. The analysis proceeds in three parts: Section \ref{subsec:3.1} establishes the dependence of the limit cycle lift, drag, and PSub displacement response on position at the highest $\lambda$ considered, revealing four distinct spatial regimes amplifying or attenuating the lift and drag oscillations; Section \ref{subsec:3.2} introduces the characteristic frequency and phase difference between PSub displacement and lift that map into these amplification or attenuation regimes; and Section \ref{subsec:3.3} examines the oscillating and time-averaged mean flow fields to understand the physical mechanisms underlying each regime.
\subsection{Simulation results for FSI dependence on PSub position}
\begin{figure}[H]
    \centering
    \includegraphics[width=0.9\textwidth]{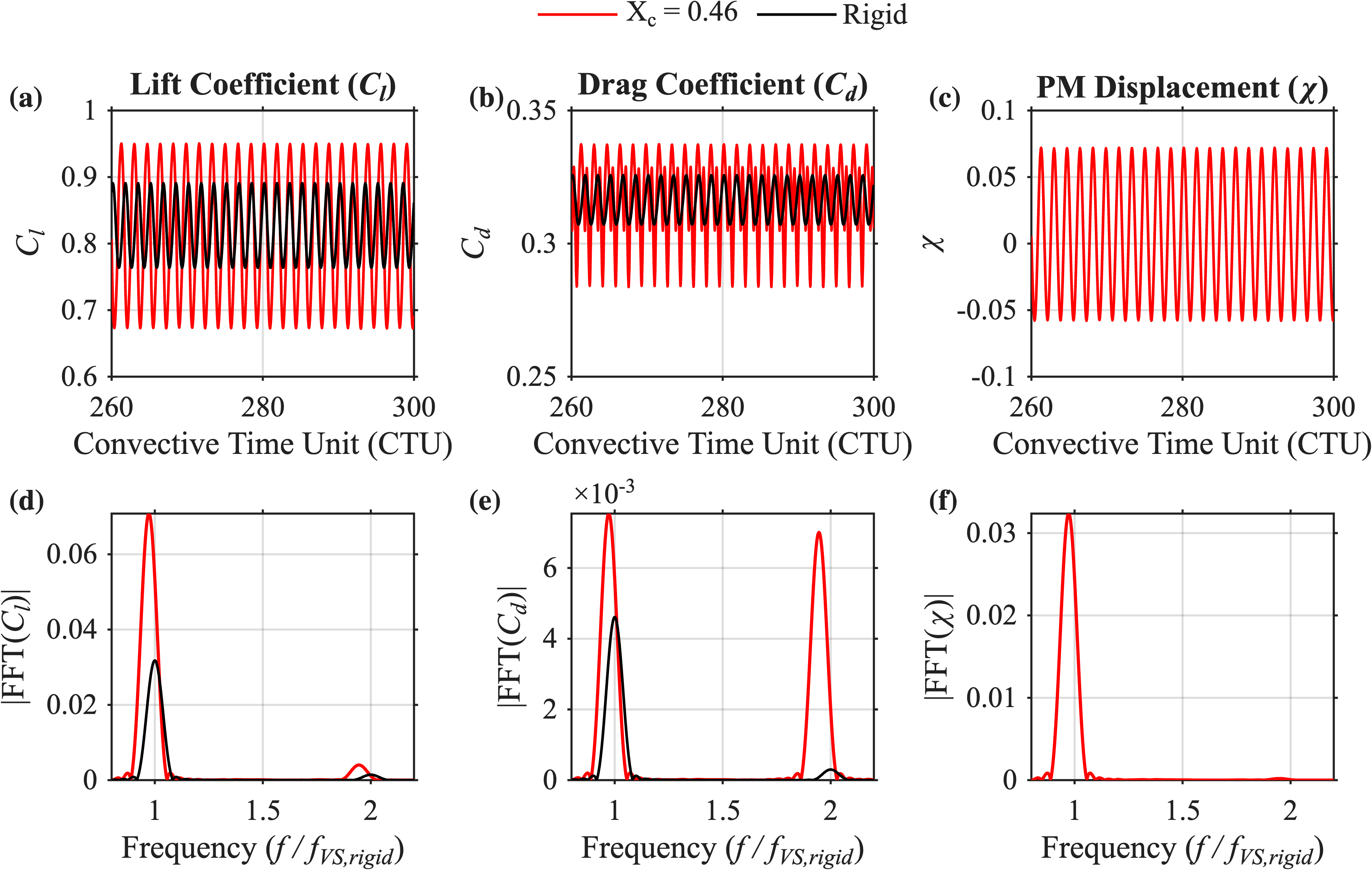}
    \caption{Representative results for $X_c = 0.46$, a case with an amplification of the oscillation amplitude of lift and drag in the limit cycle: Time histories (top row) and frequency spectra (bottom row) of the lift coefficient $C_l$ (left), drag coefficient $C_d$ (center), and PSub displacement oscillation $\chi$ (right) for $X_c = 0.46$ compared to the rigid baseline.}

    \label{fig:timefft46}
\end{figure}

\begin{figure}[H]
    \centering
    \includegraphics[width=0.9\textwidth]{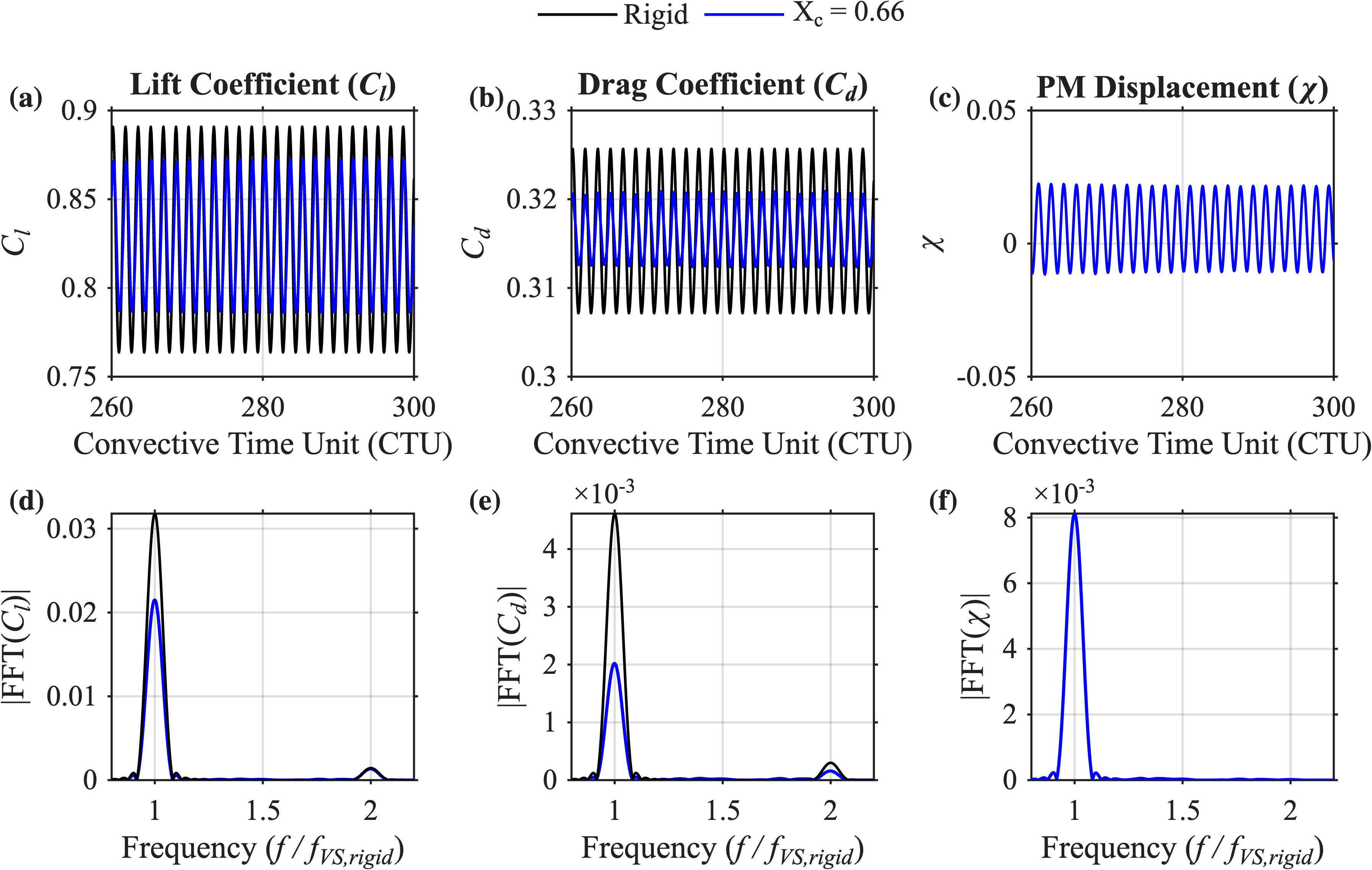}
    \caption{Representative results for $X_c = 0.66$, a case with an attenuation of the oscillation amplitudes of lift and drag in the limit cycle: Time histories (top row) and frequency spectra (bottom row) of the lift coefficient $C_l$ (left), drag coefficient $C_d$ (center), and PSub displacement oscillation $\chi$ (right) for $X_c = 0.46$ compared to the rigid baseline.}

    \label{fig:timefft66}
\end{figure}

We first show in detail the effect of the PSub on FSI quantities for two representative positions, $X_c = 0.46$ and $X_c = 0.66$, that show very different effects on lift and drag. We then present the full set of results of the flow quantities as a function of chordwise position. Figs.~\ref{fig:timefft46} and ~\ref{fig:timefft66} show the limit cycle time histories from 260 to 300 convective time units, capturing 60 oscillation cycles. The frequency spectra were computed through Fast Fourier Transform (FFT) within this time window, using zero padding to the next power of two exceeding ten times the original time window, for the lift coefficient $C_l$, drag coefficient $C_d$, and PSub displacement $\chi$ at two chordwise positions, $X_c = 0.46$ and $X_c = 0.66$, compared to the rigid baseline case. The results for both $C_l$ and $C_d$ show that when the PSub is at $X_c = 0.46$, the amplitudes of both flow quantities are significantly amplified compared to the rigid case, but when the PSub is at $X_c = 0.66$, the amplitudes of both flow quantities are significantly attenuated compared to the rigid case. Throughout this paper, for any signal $\psi(t)$ evaluated at the limit cycle (where $\psi$ represents $C_l$, $C_d$, or the PSub displacement $\chi$), the oscillation amplitude $A_\psi$ and the mean $\bar{\psi}$ are defined as
\begin{equation}
    A_\psi = \frac{\psi_{\max} - \psi_{\min}}{2}, \qquad \bar{\psi} = \frac{1}{T}\int_{t_0}^{t_0+T} \psi(t)\, dt,
    \label{eq:amp_mean}
\end{equation}
where $\psi_{\max}$ and $\psi_{\min}$ are the maximum and minimum of $\psi(t)$ over one limit cycle period $T$. Unless otherwise stated, the oscillation amplitude and mean of $C_l$, $C_d$, and $\chi$ refer to these limit cycle values, denoted $A_{C_l}, A_{C_d}, A_\chi$ and $\bar{C}_L, \bar{C}_D, \bar{\chi}$. We refer to a PSub position that amplifies the flow quantities relative to the rigid case as an ``amplification case'' and a PSub position that attenuates these flow quantities relative to the rigid case as an ``attenuation case''.  The amplification case (Fig.~\ref{fig:timefft46}) exhibits significantly larger oscillation amplitudes in both lift and drag that increase with time before reaching a steady-state amplitude, while the rigid baseline case shows much shorter initial transient before settling onto the limit cycle amplitudes of lift and drag. The dominant frequency for lift and drag shifts downward compared to the rigid case. The attenuation case  (Fig.~\ref{fig:timefft66}) shows attenuated oscillation amplitudes at the characteristic frequency of the rigid baseline after some initial transients and low-frequency beating.  The amplitude of the PSub displacement oscillation also significantly depends on the PSub position: at $X_c = 0.46$, the PSub amplitude maintains large persistent oscillations throughout the simulation, whereas at $X_c = 0.66$ the amplitude of the displacement oscillation settles to a smaller amplitude after an initial transient with low-frequency beating. 

\begin{figure}[H]
    \centering
    \includegraphics[width=0.9\textwidth]{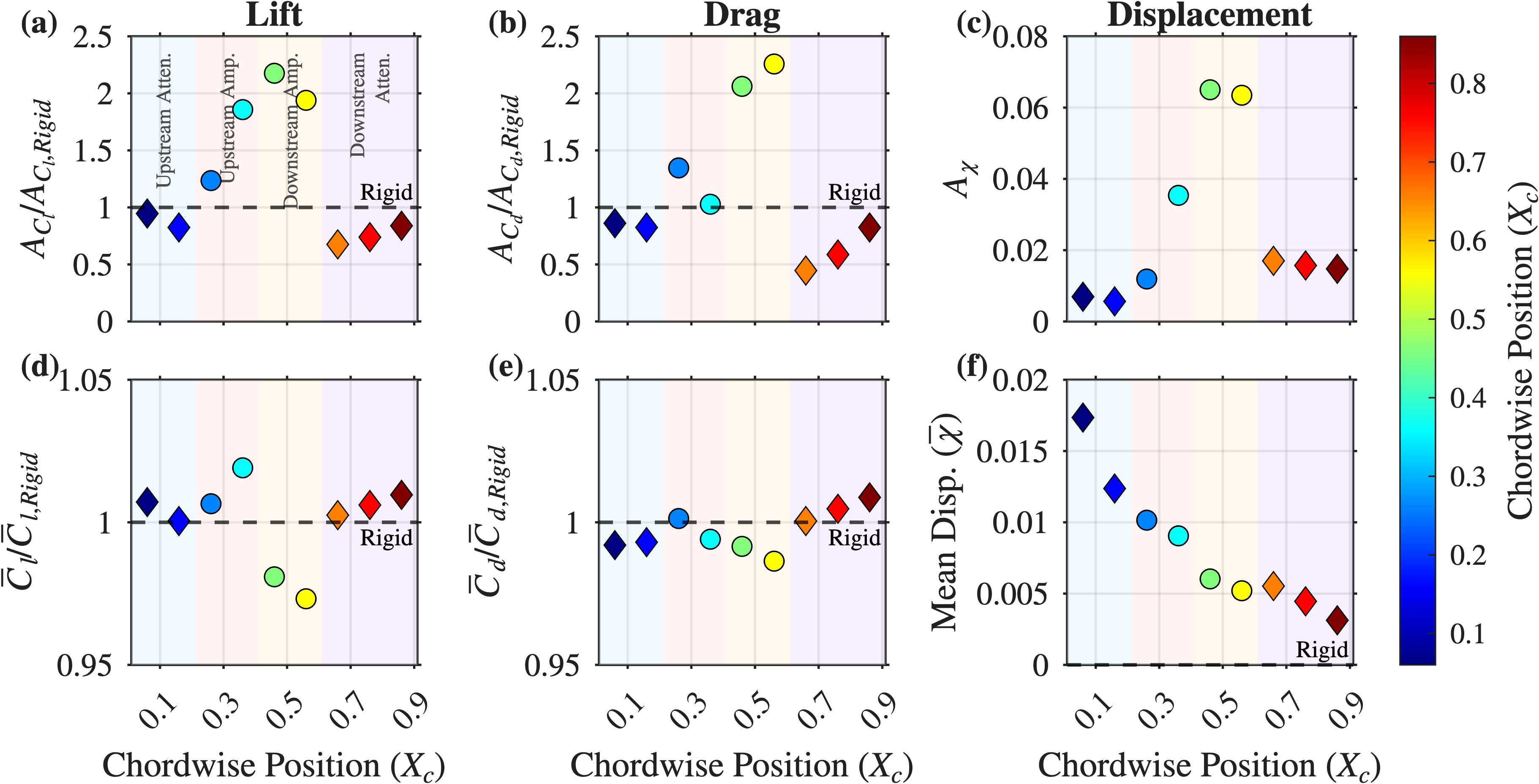}
    \caption{
    Dependence of limit cycle lift, drag, and PSub displacement on PSub chordwise position for: (a) lift amplitude, (b) drag amplitude, (c) PSub displacement amplitude, (d) mean lift, (e) mean drag, and (f) mean PSub displacement. All aerodynamic amplitudes and mean values are normalized by the rigid case. Circles and diamond markers distinguish amplification and attenuation cases, respectively. Four spatial regimes are identified (color-shaded): upstream attenuation ($X_c \leq 0.16$), upstream amplification ($X_c = 0.26, 0.36$), downstream amplification ($X_c = 0.46, 0.56$), and downstream attenuation ($X_c \geq 0.66$). The colorbar shows $X_c$, consistent with later figures.}
    \label{fig:amplitude}
\end{figure}

Figure~\ref{fig:amplitude} shows the synthesized simulation results of the flow quantities and PM displacement as a function of chordwise position. Specifically, panels (a)--(c) of Fig.~\ref{fig:amplitude} show the amplitude in the limit cycle of lift and drag normalized by the rigid case, and the amplitude of the PSub displacement in the limit cycle, while panels (d)--(f) show the corresponding mean lift, drag, and displacement, all as a function of $X_c$. The results reveal that the oscillation amplitudes depend strongly on the PSub position, while the mean flow quantities change less than the oscillation amplitudes and are generally within 3\% of the rigid baseline values.

The lift oscillation amplitude in the limit cycle (Fig.~\ref{fig:amplitude}(a)) peaks when $X_c = 0.46$, while it is most attenuated when $X_c = 0.66$. The drag oscillation amplitude (Fig.~\ref{fig:amplitude}(b)) shows the peak when $X_c = 0.56$ and most attenuated when $X_c = 0.66$. Despite this slight difference in the peak position, every position that amplifies the lift oscillation also amplifies the drag oscillation, and every position that attenuates one attenuates the other. It is also worth noting the sharp transition from amplification to attenuation between $X_c = 0.56$ and $0.66$, suggesting the PM--FSI is highly sensitive around this range of chordwise position, which is further investigated in the following flow field analysis.

The oscillation amplitude of the PSub displacement (Fig.~\ref{fig:amplitude}(c)) generally follows the lift oscillation amplitude dependence on chordwise position in the amplification regime ($X_c = 0.36$--$0.56$). We note that in the upstream positions of $X_c = 0.16$ and $X_c = 0.26$, the PSubs have nearly identical displacement amplitudes, yet $X_c = 0.16$ results in attenuation and $X_c = 0.26$ results in amplification. This shows that the limit cycle vortex shedding is sensitive to the PSub's chordwise position relative to the flat plate and the flow structure, and not simply to the amplitude of the PSub displacement; the timescales and dynamics of that displacement, such as its frequency and phase relative to the flow, could also play a role. At downstream positions, the displacement can become negative during oscillation cycles, meaning the compliant section intrudes into the pressure side at portions of the oscillation. This intermittent pressure side intrusion at the downstream positions due to the small mean displacement and large oscillation amplitude of the displacement, particularly $X_c \geq 0.46$, will be discussed further in the flow field analysis (Section \ref{subsec:3.3}). 

The mean lift and drag do not necessarily change in the same direction as the oscillation amplitudes. The mean lift coefficient (Fig.~\ref{fig:amplitude} (d)) shows a slight increase at the upstream amplification positions ($X_c = 0.26, 0.36$) and a slight decrease at the downstream amplification positions ($X_c = 0.46, 0.56$). The mean drag coefficient (Fig.~\ref{fig:amplitude}(e)) shows a slight decrease at the upstream attenuation positions ($X_c = 0.06, 0.16$) and a slight increase at the downstream attenuation positions ($X_c = 0.66, 0.76, 0.86$). The changes in the mean lift and drag are less pronounced than the oscillation amplitudes. The PSub mean displacement (Fig.~\ref{fig:amplitude}(f)) generally decreases from upstream to downstream positions, with the exception of the transition from $X_c = 0.66$ to $0.76$.

\begin{figure}[H]
    \centering
    \includegraphics[width=0.9\textwidth]{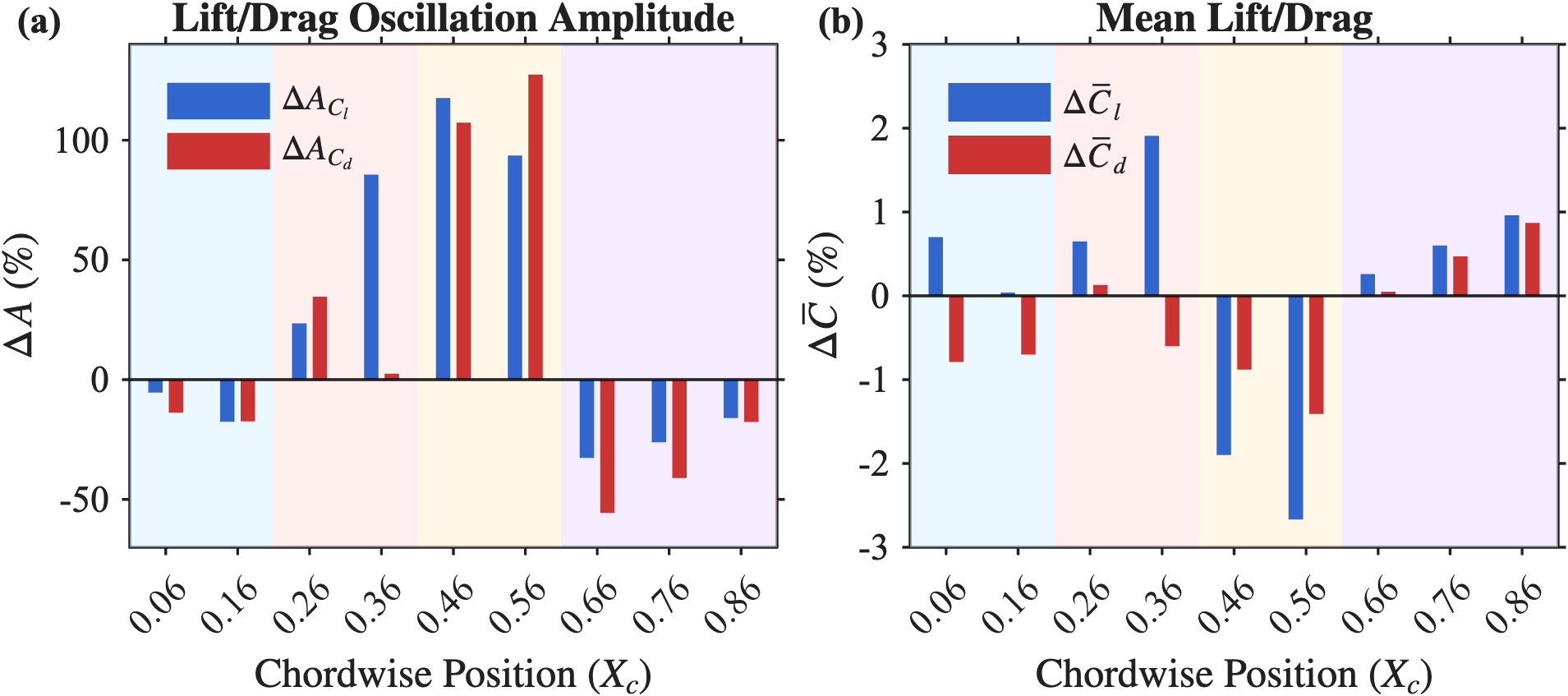}
    \caption{Percentage changes relative to the rigid baseline at each chordwise position $X_c$ for (a) the lift and drag oscillation amplitudes, $\Delta A_{C_l}$ and $\Delta A_{C_d}$, and (b) the mean lift and drag, $\Delta \bar{C}_l$ and $\Delta \bar{C}_d$. Background shading indicates the four spatial regimes, with colors corresponding to the regime colors in Fig.~\ref{fig:amplitude}.}
    \label{fig:regimes}
\end{figure}

These lift and drag dependencies on the chordwise position in Figure~\ref{fig:amplitude} motivate the identification of distinct spatial regimes for characterizing distinct FSI mechanisms. Figure~\ref{fig:regimes} summarizes the precentage changes relative to the rigid baseline in the limit cycle oscillation amplitudes We define these regimes using two criteria in a sequence. The first criterion is the whether the oscillation amplitude increases or decreases relative to the rigid baseline. This splits the nine positions into an amplification regime ($X_c = 0.26$--$0.56$, where $A_{C_l}/A_{C_l,\mathrm{rigid}} > 1$) and an attenuation regime ($X_c \leq 0.16$ and $X_c \geq 0.66$, where $A_{C_l}/A_{C_l,\mathrm{rigid}} < 1$).

In addition to these two regimes identified using the oscillation amplitude, we can also take into account the mean lift and drag changes. We use mean lift or mean drag as the second criterion, identifying the amplification and attenuation regimes above into an upstream and a downstream sub-regime, for four regimes in total. Within the amplification regime, the mean lift increase or decrease separates the upstream positions from downstream positions: $X_c = 0.26, 0.36$ showing a small increase in mean lift and $X_c = 0.46, 0.56$, showing a decrease. Within the attenuation regime, the sign of the mean drag change separates the upstream positions ($X_c = 0.06, 0.16$), which show a small decrease in mean drag, from the downstream positions ($X_c = 0.66, 0.76, 0.86$), which show an increase.

The attenuation of the lift and drag oscillations at $X_c = 0.66$ is particularly noteworthy as it achieves approximately 30\% reduction in lift oscillation and 50\% reduction in drag oscillation, while causing less than 0.5\% change in mean lift and mean drag. This particular result is significant as it suggests PSubs at particular positions could passively mitigate unsteadiness without detrimentally affecting the mean flow quantities of lift and drag.
\label{subsec:3.1}

\subsection{Dependence of Frequency and Phase on PSub Position}
\begin{figure}[H]
    \centering
    \includegraphics[width=0.9\textwidth]{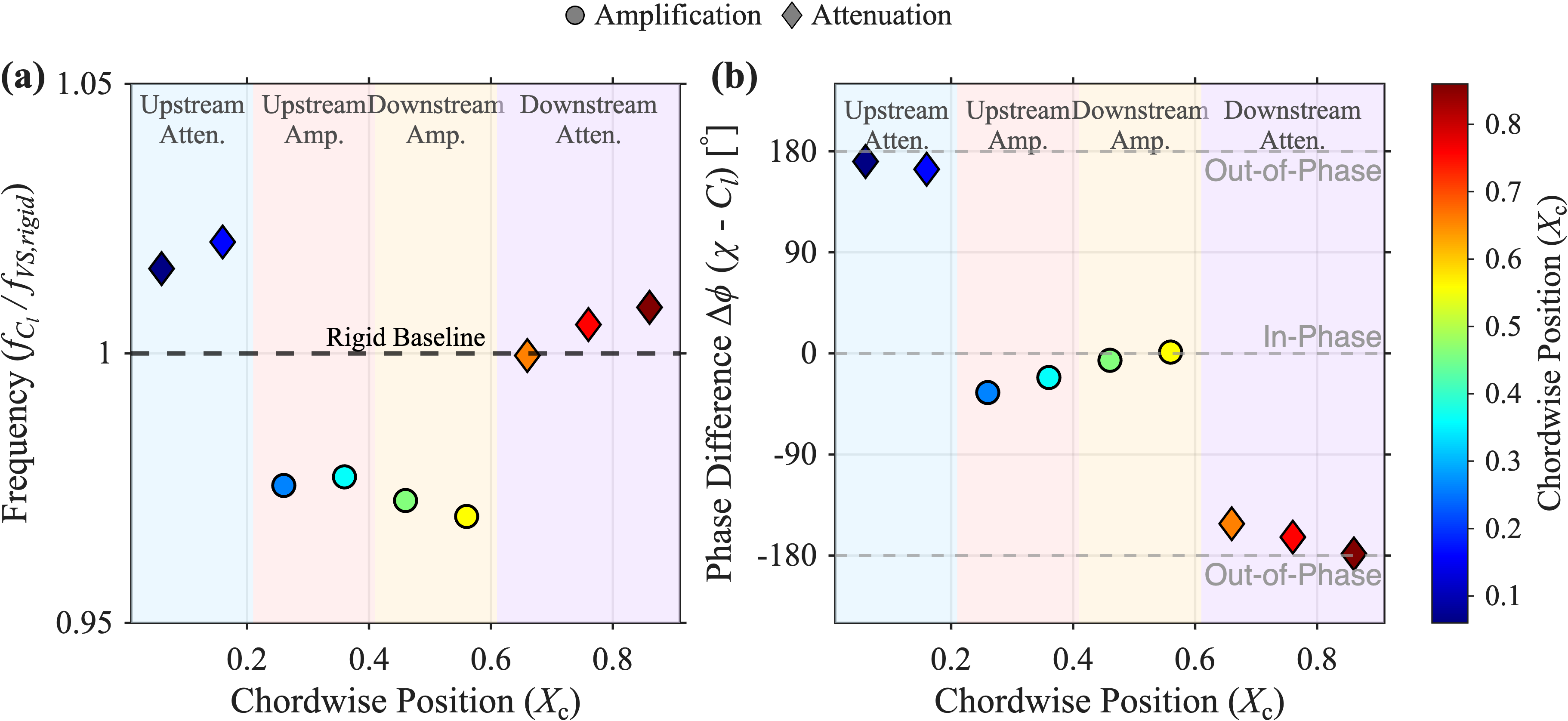}
    \caption{Characteristic frequency $f_{C_l}/f_{\mathrm{rigid}}$ (left) and phase difference $\Delta\phi(\chi, C_l)$ (right) as a function of PSub chordwise position $X_c$, with the four spatial regimes labeled. Amplification PSub positions (0.26-0.56) show lower characteristic frequencies ($>2\%$ below rigid) and in-phase relationship between displacement and lift, while leading edge and trailing edge positions ($X_c \le 0.16, X_c\ge0.66$) show near-rigid or higher characteristic frequencies and out-of-phase relationship between displacement and lift.}

    \label{fig:freq}
\end{figure}

With the four distinct chordwise regions identified above, we now examine the characteristic frequency in the lift signal $f_{C_l}/f_{\mathrm{rigid}}$, and the phase difference, $\Delta\phi(\chi, C_l)$, at that frequency between PSub displacement and lift, as a function of the chordwise position (Figure~\ref{fig:freq}). These quantities are evaluated through FFT at the limit cycle from 260 to 300 convective time units, spanning more than 20 cycles of the lift and displacement oscillation. We only present the lift results here for brevity, as the drag shows the same dependency of characteristic frequency and phase on the chordwise position as the lift. Results show that the frequency and phase depend strongly on whether the PSub position is in the amplification or attenuation regime. All amplification positions ($X_c = 0.26$--$0.56$) exhibit characteristic frequencies more than 2\% below the rigid baseline and a near-zero phase difference, indicating that the PSub displacement and lift oscillations are in-phase. In contrast, all attenuation positions ($X_c \leq 0.16$ and $X_c \geq 0.66$) maintain characteristic frequencies near or slightly above the rigid baseline, with an out-of-phase relationship ($\Delta\phi \approx \pm 180^\circ$) between displacement and lift.

\begin{figure}[H]
    \centering
    \includegraphics[width=0.9\textwidth]{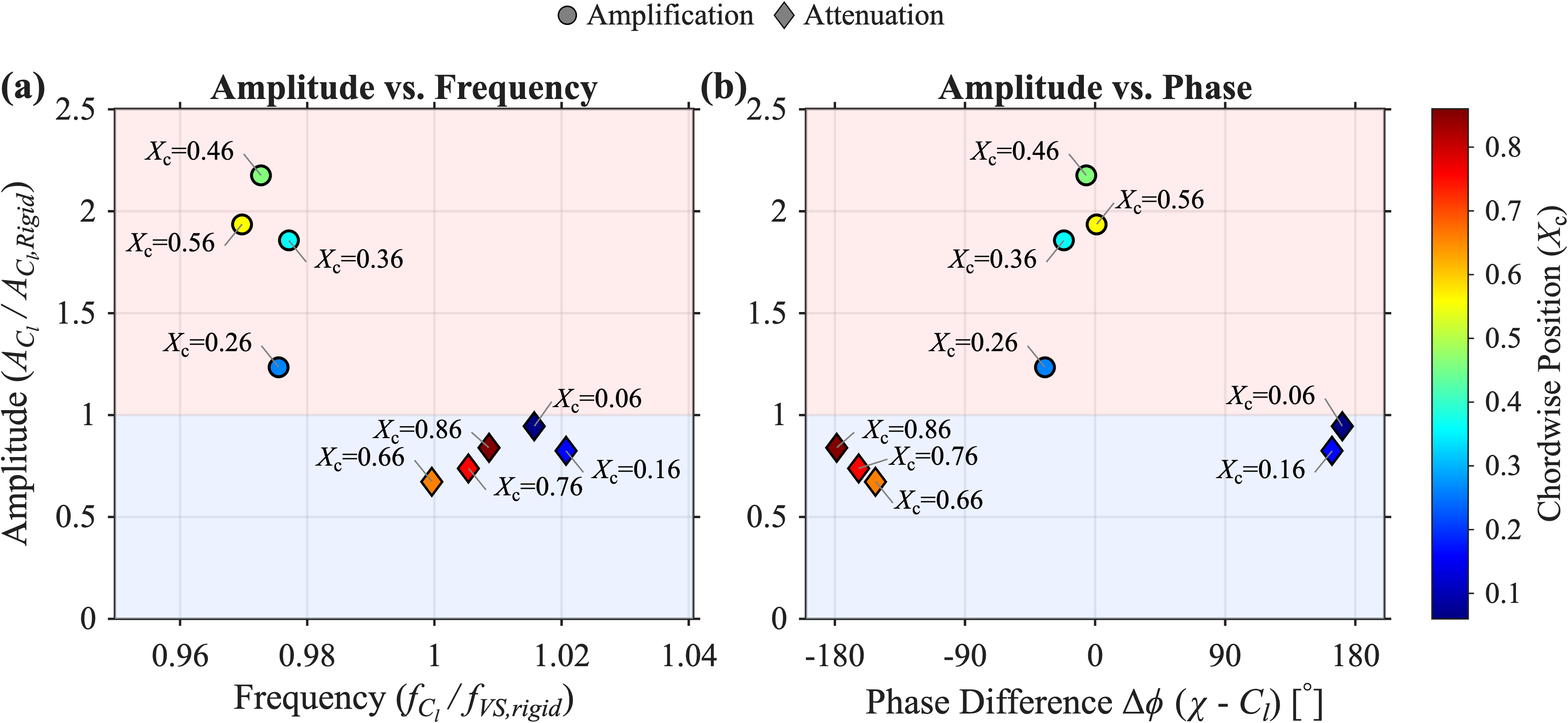}
    \caption{Lift oscillation $A_{C_l}/A_{C_l,\mathrm{rigid}}$ vs. normalized lift oscillation frequency $f_{C_l}/f_{\mathrm{rigid}}$ (left) and phase difference $\Delta\phi$ between PSub displacement $\chi$ and lift coefficient $C_l$ (right), colored by chordwise position $X_c$. Amplification ($A_{C_l}/A_{C_l,\mathrm{rigid}} > 1$, circles) and attenuation ($A_{C_l}/A_{C_l,\mathrm{rigid}} < 1$, diamonds) cases are distinguished by marker shape. Dashed lines indicate rigid baseline values.}
    \label{fig:binary}
\end{figure}

The relationship between the lift amplitude and the lift frequency or phase differences between the lift and displacement is examined in Figure~\ref{fig:binary}. In the amplification cases, the lift signal responds at a frequency  2\%  lower than the rigid vortex-shedding frequency, and responds in-phase with the PSub displacement at this frequency. On the other hand, in the attenuation cases, the oscillation frequency ranges from the nearly the same frequency as the rigid case or increases up to 2\%, but the lift and PSub displacement responses are approximately out-of-phase.

It is also important to note that all four amplification positions share a nearly identical in-phase relationship, despite spanning a range of lift amplitudes from 20\% to 120\% above the rigid baseline. Similarly, all attenuation positions maintain a similar out-of-phase relationship regardless of whether the attenuation in lift oscillation amplitude is small (e.g., $\sim 5\%$ reduction for $X_c = 0.06$) or significant (e.g., $\sim 30\%$ reduction for $X_c = 0.66$). This suggests a phase lock between the PSub and flow at the limit cycle, that depends  on the PSub's position relative to the flow structure rather than by the magnitude of the oscillation. This spatial relationship, specifically the PSub's position relative to the unsteady recirculation, is further examined and quantified in Section~\ref{subsec:3.3}.

One possible interpretation of the relation between the frequency and phase shift is that when the PSub displacement and lift oscillations are in-phase, this effectively increases the mass participating in the oscillation cycle, which reduces the characteristic frequency. In contrast, the out-of-phase motion of the PSub relative to the lift oscillation cycle in attenuation cases may act as an effective added stiffness opposing the forcing from the flow, increasing the vortex-shedding frequency. While a definitive explanation would require a detailed analysis of the added mass and fluid-structure coupling at each position, the observed relationships are qualitatively consistent with vortex-induced vibration behavior, where the phase relationship between structural motion and flow forcing is known to influence the effective oscillation frequency \citep{williamson2004vortex, sarpkaya2004critical}.

\label{subsec:3.2}

\subsection{Flow Field Analysis and Discussion}
\label{subsec:3.3}
To understand the physical mechanisms underlying the four regimes, we examine the unsteady and time-averaged flow velocity and vorticity fields for representative positions: $X_c = 0.16$ (upstream attenuation), $X_c = 0.36$ (upstream amplification), $X_c = 0.46$ (downstream amplification), and $X_c = 0.66$ (downstream attenuation). This analysis is given below in two sections: Section \ref{subsubsec:3.3.1} examines the root mean square (RMS) velocity and vorticity fields to identify how each PSub position alters the unsteady oscillating part of the flow, while Section \ref{subsubsec:3.3.2} examines the mean vorticity and flow velocity changes, which correlate with changes to the mean $C_l$ and $C_d$.

\subsubsection{PSub-induced Modifications to the Unsteady Flow}
\label{subsubsec:3.3.1}

\begin{figure}[H]
    \centering
    \includegraphics[width=0.9\textwidth]{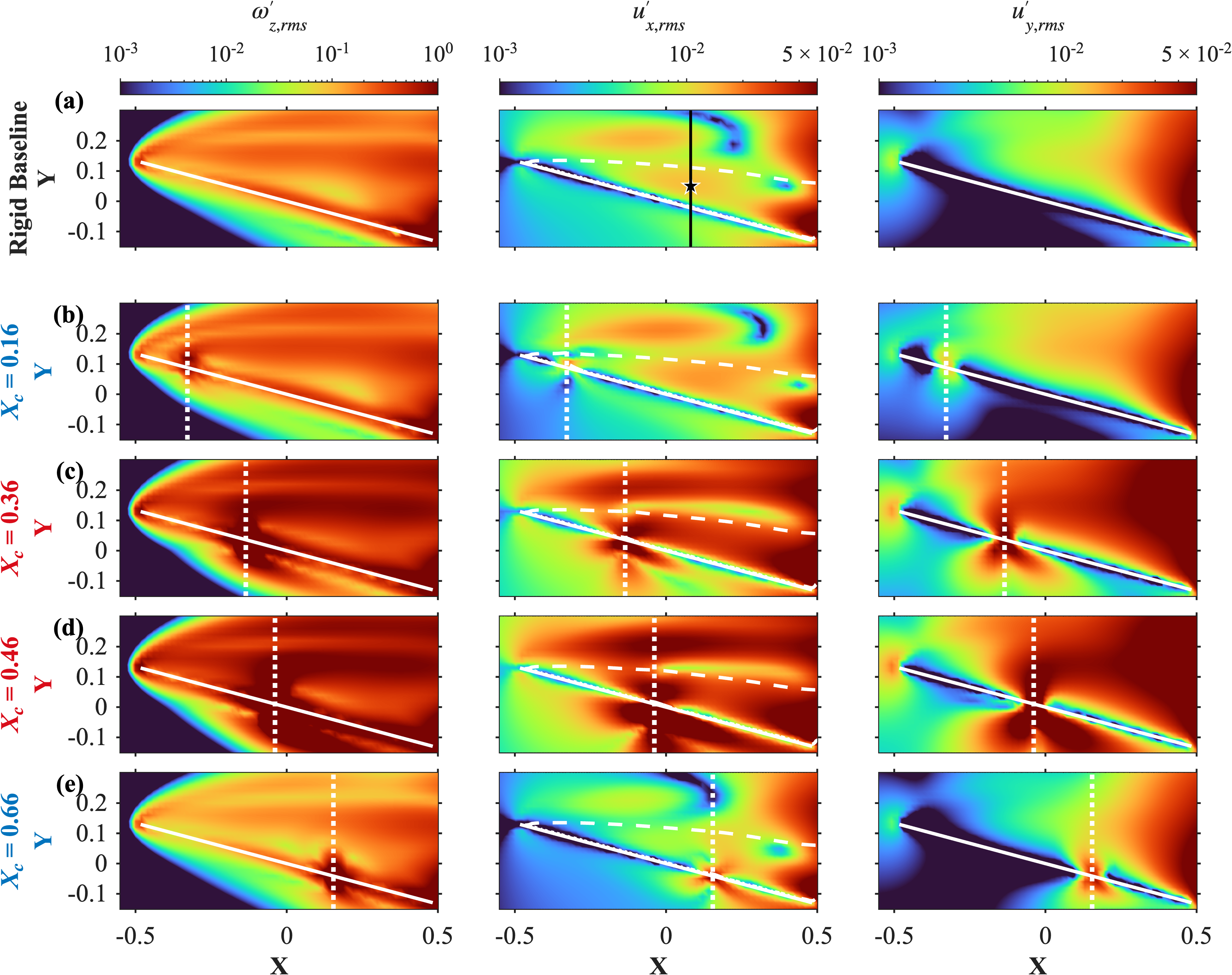}
    \caption{RMS fields of vorticity $\omega'_{z,{rms}}$ (left), streamwise velocity $u'_{x,{rms}}$ (center), and transverse velocity $u'_{y,{rms}}$ (right) for the rigid baseline (top row) and PSub cases at $X_c = 0.16, 0.36, 0.46, 0.66$ (rows 2--5), shown on a logarithmic color scale. The dashed white contour indicates the $\bar{u}_x = 0$ boundary for each case, marking the recirculation zone. The vertical black solid line with the star marker illustrates the peak $u'_{x,rms}$ ($X_{c,\mathrm{peak}}=0.58$) inside the recirculation zone for the rigid case. The vertical white dotted lines in rows 2--5 show the X position of the center of the PSub.}
    \label{fig:rms_fields}
\end{figure}

We first analyze the unsteady flow fields for the rigid baseline case to which we will compare the PSub-induced flow fields. For any flow variable $\phi(\mathbf{x}, t)$, the RMS and time-average flow field variables are defined as:
\begin{equation}
    \phi'_{\mathrm{rms}}(\mathbf{x}) = \sqrt{\frac{1}{T} \int_{t_0}^{t_0 + T} \left[\phi(\mathbf{x}, t) - \bar{\phi}(\mathbf{x})\right]^2 \, dt}, \qquad \bar{\phi}(\mathbf{x}) = \frac{1}{T} \int_{t_0}^{t_0 + T} \phi(\mathbf{x}, t) \, dt,
    \label{eq:time_avg_rms}
\end{equation}
The rigid baseline RMS vorticity and velocity fields (top row of Fig.~\ref{fig:rms_fields}) show the spatial characteristics of the intrinsic vortex-shedding process. The RMS vorticity, $\omega'_{z,\mathrm{rms}}$, illustrates the oscillation of the vorticity near the plate, which can show the degree of unsteadiness in the flow due to the vortex shedding. The RMS streamwise velocity, $u'_{x,rms}$, field shows the oscillation of the streamwise velocity and highlights two distinct high-oscillation regions. The first region is between the freestream and the $\bar{u}_x = 0$ boundary (dashed white contour), corresponding to the oscillation of the separated shear layer. The second region is below the $\bar{u}_x = 0$ boundary and closer to the plate surface, where periodic reverse flow occurs within a recirculation zone. Within this recirculation zone, the location of the peak RMS streamwise velocity, defined as $X_{c,\mathrm{peak}}$, is marked with a black star in the $u'_{x,\mathrm{rms}}$ column of Fig.~\ref{fig:rms_fields} row (a). This location corresponds to the chordwise position $X_{c,\mathrm{peak}} = 0.58$, and is indicated with a black vertical solid line so that it can be related to the PSub positions shown in the subsequent rows. We refer to this peak as the near-surface $u'_{x,rms}$ peak. The RMS transverse velocity $u'_{y,rms}$ field shows oscillations which are more prominent on the suction side, where the shear layer and the recirculation region exist.

The PSub cases (rows 2--5 of Fig.~\ref{fig:rms_fields}) show distinct changes to this baseline flow structure. At the upstream attenuation position $X_c = 0.16$, the lift and drag oscillations on the plate are reduced relative to the rigid baseline (Figs.~\ref{fig:amplitude}--\ref{fig:regimes}). Compared to the rigid case, $\omega'_{z,\rm rms}$ shows a slight reduction right at the leading edge but increases near the PSub and along the suction side. $u'_{x,\rm rms}$ shows a similar pattern: a slight reduction at the leading edge on the suction side, but increased intensity at the shear layer and the near-surface $u'_{x,\rm rms}$ peak in the recirculation zone, with a slight downstream shift. Because the PSub at this position lies near the onset of recirculation and remains distant from the primary near-surface $u'_{x,\rm rms}$ peak, its effect on these instabilities is limited to slight, localized changes near the leading edge and the PSub, without altering the broader flow structure.

At the upstream amplification position $X_c = 0.36$, all three $\omega'_{z,rms}, u'_{x,rms}, u'_{y,rms}$ fields show increased oscillations on the suction side. The PSub at this position sits just upstream of the near-surface $u'_{x,rms}$ peak within the recirculation, where the perturbations introduced by the PSub motion are carried downstream into the near-surface $u'_{x,rms}$ peak. Combined with the in-phase relationship between the PSub displacement and the lift oscillation (Section \ref{subsec:3.2}), the lift and drag oscillation amplitudes increase at this position. It is noted that the high-intensity oscillation in all three $\omega'_{z,rms}, u'_{x,rms}, u'_{y,rms}$ fields near the PSub on the pressure side does not fully extend to the trailing edge. The limited pressure side impact at $X_c = 0.36$ is consistent with the higher mean displacement and lower displacement oscillation observed at this position.

At the downstream amplification position $X_c = 0.46$, the $\omega'_{z,rms}, u'_{x,rms}, u'_{y,rms}$ fields show increased oscillations on both the suction and pressure sides. Unlike $X_c = 0.36$, where the modifications were dominant on the suction side, $X_c = 0.46$ disrupts the flow on both sides of the plate. As discussed in Section \ref{subsec:3.1}, the downstream PSub positions have smaller mean displacement and the displacement can become negative during portions of the cycle, intruding more into the pressure side flow. This disruption on both suction and pressure sides observed at $X_c = 0.46$ coincides with the larger amplification of drag oscillation, in contrast to the suction-side-dominant modification at $X_c = 0.36$ where drag oscillation remains close to that of the rigid case.

At the downstream attenuation position $X_c = 0.66$, the $\omega'_{z,rms}, u'_{x,rms}, u'_{y,rms}$ fields show attenuated oscillations on the suction side and locally increased oscillations on the pressure side near the trailing edge. The unsteadiness within the recirculation region and the shear layer is noticeably weaker than the rigid baseline. The PSub at this position sits downstream of the near-surface $u'_{x,rms}$ peak, where the near-surface and shear layer $u'_{x,rms}$ have begun to weaken. Unlike the in-phase PSub motion at $X_c = 0.36$ and $0.46$, where the PSub sits just upstream of the near-surface $u'_{x,rms}$ peak and amplifies it, the PSub at $X_c = 0.66$ sits downstream of the peak, receives the vortex shedding from upstream, and opposes the lift oscillation through its out-of-phase motion. While the suction side oscillations are broadly attenuated, a localized increase in $u'_{x,\rm rms}$ appears from the PSub extending to the trailing edge on the pressure side. This may suggest that the PSub not only interferes with the suction side recirculation but also interacts with the trailing-edge vortex shedding, possibly disrupting the coherent vortex formation in the wake and resulting in weaker lift and drag oscillations due to the vortex shedding.

These results can be correlated with the sharp transition from large amplification to large attenuation in the oscillation amplitudes of lift and drag between $X_c=0.56$ and $0.66$ in Figs.~\ref{fig:amplitude} and \ref{fig:regimes}: positions just upstream of the peak amplify the oscillations, consistent with $X_c = 0.36$ and $0.46$ above, while positions downstream of the peak consistently attenuate them, consistent with $X_c = 0.66$ above. However, the upstream attenuation positions ($X_c = 0.06, 0.16$) fall on the same side of the peak as the amplification cases, yet still attenuate the oscillations. This suggests that these upstream attenuation positions ($X_c = 0.06, 0.16$) interact with the surrounding flow through different mechanisms, which is further investigated in the following sections.

\subsubsection{PSub-induced Modifications to the Mean Flow}
\label{subsubsec:3.3.2}
\begin{figure}[H]
    \centering
    \includegraphics[width=0.9\textwidth]{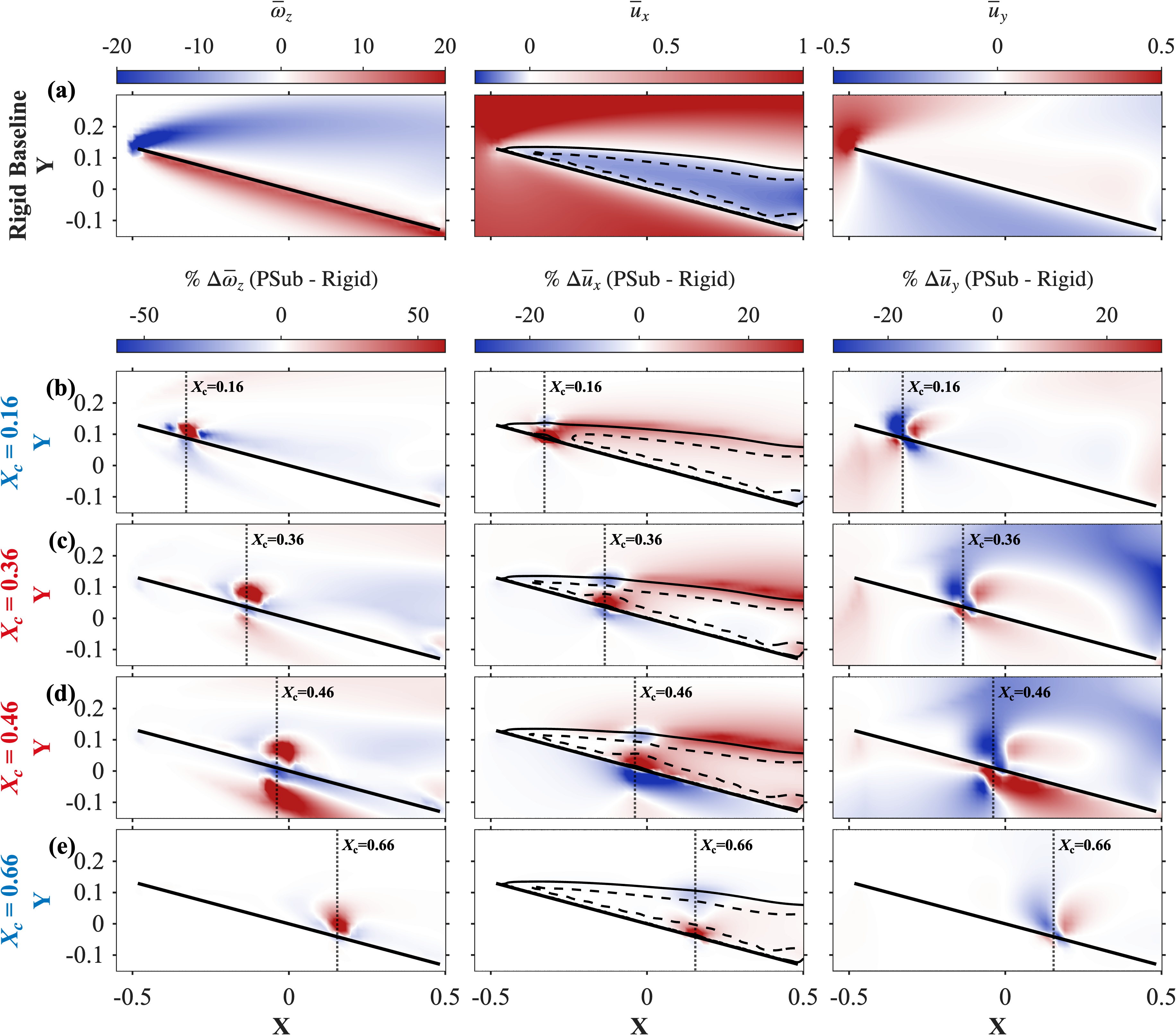}
    \caption{ (a) Mean flow fields for the rigid baseline and (b-e) percentage difference from rigid ($\%\Delta = 100 \times (\bar{\square}_{\mathrm{PSub}} - \bar{\square}_{\mathrm{rigid}}) / \max|\bar{\square}_{\mathrm{rigid}}|$) for each PSub case at (b) $X_c = 0.16$ (c) $X_c = 0.36$, (d)$X_c = 0.46$, (e) $X_c = 0.66$: vorticity $\bar{\omega}_z$ (left), streamwise velocity $\bar{u}_x$ (center), and vertical velocity $\bar{u}_y$ (right). The $\bar{u}_x = 0$ (solid) and $\bar{u}_x = -0.08$ (dashed) contour lines are overlaid on the center column showing the mean streamwise velocity ($\bar{u}_x$) for each case.}
    \label{fig:mean_fields_diff}
\end{figure}

We next analyze the mean flow fields (Fig.~\ref{fig:mean_fields_diff}) for velocity and vorticity, focusing first on understanding the flow fields for the rigid baseline with which to compare the configurations with a PSub. The rigid baseline mean flow (top row of Fig.~\ref{fig:mean_fields_diff}) shows the fully separated structure characteristic of the post-bifurcation regime: a large recirculation region delineated by the $\bar{u}_x = 0$ contour (solid line) on the suction side, with a representative contour at $\bar{u}_x = -0.08$ (dashed line, approximately 40\% of the minimum mean $\bar{u}_x$ in the flow field) chosen to illustrate the downstream shift of the reverse flow structure within the recirculation. Note that in all cases, the $\bar{u}_x = 0$ contour remains close to the rigid baseline, while the $\bar{u}_x = -0.08$ contour shifts depending on the position of the PSub, with the largest change observed at $X_c=0.16$. The mean vorticity field $\bar{\omega}_z$ of the rigid baseline case shows a concentrated negative vorticity band along the suction side surface from the leading edge, marking the separated shear layer in its time-averaged form. The mean $\bar{u}_y$ field of the rigid case shows an upward velocity near the leading edge associated with flow separation and a downward velocity over the rear portion of the suction side as the flow re-orients toward the wake.

Rows (b)--(d) of Fig.~\ref{fig:mean_fields_diff} show the percentage difference in the mean flow fields between each PSub case and the rigid baseline, defined as
\begin{equation}
    \%\Delta\bar{\phi} = 100 \times \frac{\bar{\phi}_{\mathrm{PSub}} - \bar{\phi}_{\mathrm{rigid}}}{\max|\bar{\phi}_{\mathrm{rigid}}|},
    \label{eq:pct_diff}
\end{equation}
where $\bar{\phi}$ represents the mean flow variable ($\bar{\omega}_z$, $\bar{u}_x$, or $\bar{u}_y$). The $\bar{u}_x = 0$ and $\bar{u}_x = -0.08$ contours are overlaid for both the PSub and rigid cases to illustrate shifts in the recirculation structure.

At $X_c = 0.16$, the $\%\Delta\bar{u}_x$ field shows increased streamwise velocity within the suction side recirculation zone, suggesting weakened reverse flow. The $\bar{u}_x = -0.08$ contour retreated downstream relative to the rigid case, suggesting that the PSub alters the flow at the onset of the recirculation. The changes due to the PSub to the $\bar{\omega}_z$ and $\bar{u}_y$ are largely localized near the PSub and do not extend significantly downstream. These observations suggest that the PSub at $X_c = 0.16$ may shorten the chordwise extent of the recirculation region. The net effect over the plate with a potentially shorter and weaker recirculation region could result in the reduced oscillation in lift and drag, despite the increased oscillation in the surrounding flow.

At $X_c = 0.36$, the $\%\Delta\bar{\omega}_z$ field shows strong modifications on the suction side that are largely localized near the PSub. The $\%\Delta\bar{u}_x$ field shows a band of increased streamwise velocity within the recirculation originating from the PSub location, while the streamwise extent of the $\bar{u}_x = -0.08$ contour remains relatively similar to the rigid baseline. The pressure side shows no significant change in the three fields. This suction-side-dominant modification is consistent with the slightly increased mean lift observed at this position. The pressure side mean flow remaining largely intact is also consistent with the observation in Section \ref{subsec:3.1} that drag oscillations remain close to the rigid case at $X_c = 0.36$, even as the lift oscillations are amplified.

In contrast, at $X_c = 0.46$, the $\%\Delta\bar{\omega}_z$ field shows significant changes on both the suction and pressure sides. The red region below the plate represents a pressure side change that is largely absent in all other cases. The $\%\Delta\bar{u}_x$ field is consistent with this, showing a notably reduced velocity (blue) on the pressure side downstream of the PSub. The $\%\Delta\bar{u}_y$ field also shows significant changes on the pressure side compared to $X_c = 0.36$. The modification on the pressure side  at $X_c = 0.46$ may be related to the mean lift decrease at this position. Although both $X_c = 0.36$ and $X_c = 0.46$ are amplification cases, $X_c = 0.46$ produces larger displacement and is located downstream where the PSub bulge may interact more directly with the pressure side flow. The changes on both sides at $X_c = 0.46$ correlate with the heavy amplification of both lift and drag oscillations at this position.

At $X_c = 0.66$, all three mean flow fields show minimal changes relative to the rigid baseline, with modifications confined to a small region near the PSub on the suction side. These results are consistent with Fig.~\ref{fig:regimes}, where unsteady oscillations are attenuated by more than 30\% without significantly affecting the mean flow.

\section{Dependence of Transient Responsiveness and Limit Cycle Characteristics on $\lambda$}
\label{sec:lambda}

With an understanding of how the PSub chordwise position, $X_c$, affects the PM--FSI, we now examine the effect of the displacement amplitude envelope, $\lambda$, on the FSI dynamics. We sweep $\lambda$ at four representative positions selected from the four regimes identified in Section \ref{subsec:3.1}: $X_c = 0.16$ (upstream attenuation), $X_c = 0.36$ (upstream amplification), $X_c = 0.46$ (downstream amplification), and $X_c = 0.66$ (downstream attenuation). The analysis is organized into two parts: Section \ref{subsec:4.1} examines how $\lambda$ affects the limit cycle behavior across the four positions, revealing that the effect of $\lambda$ on the limit cycle depends on the PSub position; Section \ref{subsec:4.2} examines the flow fields to characterize how $\lambda$  affects the flow field.

\begin{figure}[H]
    \centering
    \includegraphics[width=0.9\textwidth]{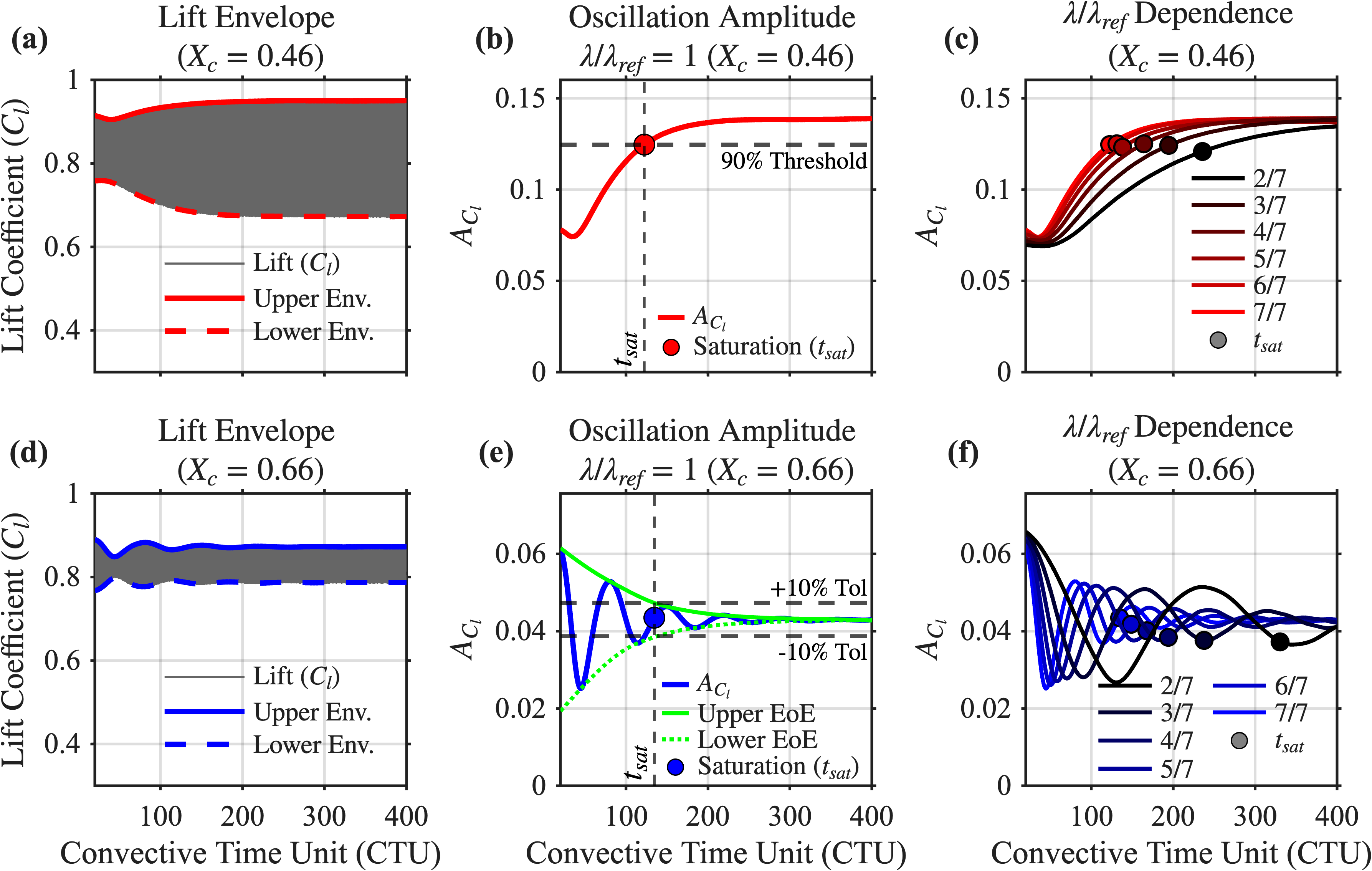}
    \caption{Lift transient analysis and $\lambda$ dependence for representative amplification ($X_c = 0.46$, (a)-(c)) and attenuation ($X_c = 0.66$, (d)-(f)) cases. (a),(d): lift coefficient time history $C_l(t)$ at $\lambda/\lambda_{\mathrm{ref}} = 1$ with extracted upper and lower envelopes (solid and dashed red/blue lines), whose half-difference defines the instantaneous oscillation $A_{C_l}(t)$. (b),(e): extracted $A_{C_l}(t)$ for the case $\lambda/\lambda_{\mathrm{ref}} = 1$, illustrating the saturation criteria. (c),(f): $A_{C_l}(t)$ across the full $\lambda$ sweep, with saturation points marked with circles that have the same color as the curve for each $\lambda$ case.}
    \label{fig:4_0}
\end{figure}

Before presenting the full dependence of lift and drag on $\lambda$ at each PSub position, we first examine the transient development of the lift oscillations to establish how the limit cycle is approached and how $\lambda$ influences this process in two representative cases. Specifically, Figure~\ref{fig:4_0} illustrates the representative amplification ($X_c = 0.46$) and attenuation ($X_c = 0.66$) cases, which are the locations of maximum amplification or attenuation of the lift oscillation amplitude, for $\lambda$ varying from $\lambda/\lambda_{\mathrm{ref}} = 2/7$ to $\lambda/\lambda_{\mathrm{ref}} = 1$. For the amplification case, the oscillation $A_{C_l}(t)$ grows monotonically after the initial transient and smoothly reaches its limit cycle value, and $t_{\mathrm{sat}}$ is defined as the time $A_{C_l}$ first reaches 90\% of its limit cycle amplitude. For the attenuation case, $A_{C_l}(t)$ exhibits a more complex transient with low-frequency beating before settling. An envelope-of-envelopes (EoE) is fitted to illustrate this attenuation, and $t_{\mathrm{sat}}$ is defined as the time the EoE enters a $\pm 10\%$ band around the limit cycle amplitude. In both cases, all $\lambda$ values converge to nearly identical limit cycle amplitudes, but they differ in that cases with higher $\lambda$ reach saturation ($t_{sat}$) earlier. These results suggest that $\lambda$ primarily affects the transient responsiveness, while $X_c$ largely determines the limit cycle lift and drag oscillation amplitudes.

\subsection{$\lambda$ Dictates the Rate to Reach the Limit Cycle}
\label{subsec:4.1}

\begin{figure}[H]
    \centering
    \includegraphics[width=0.9\textwidth]{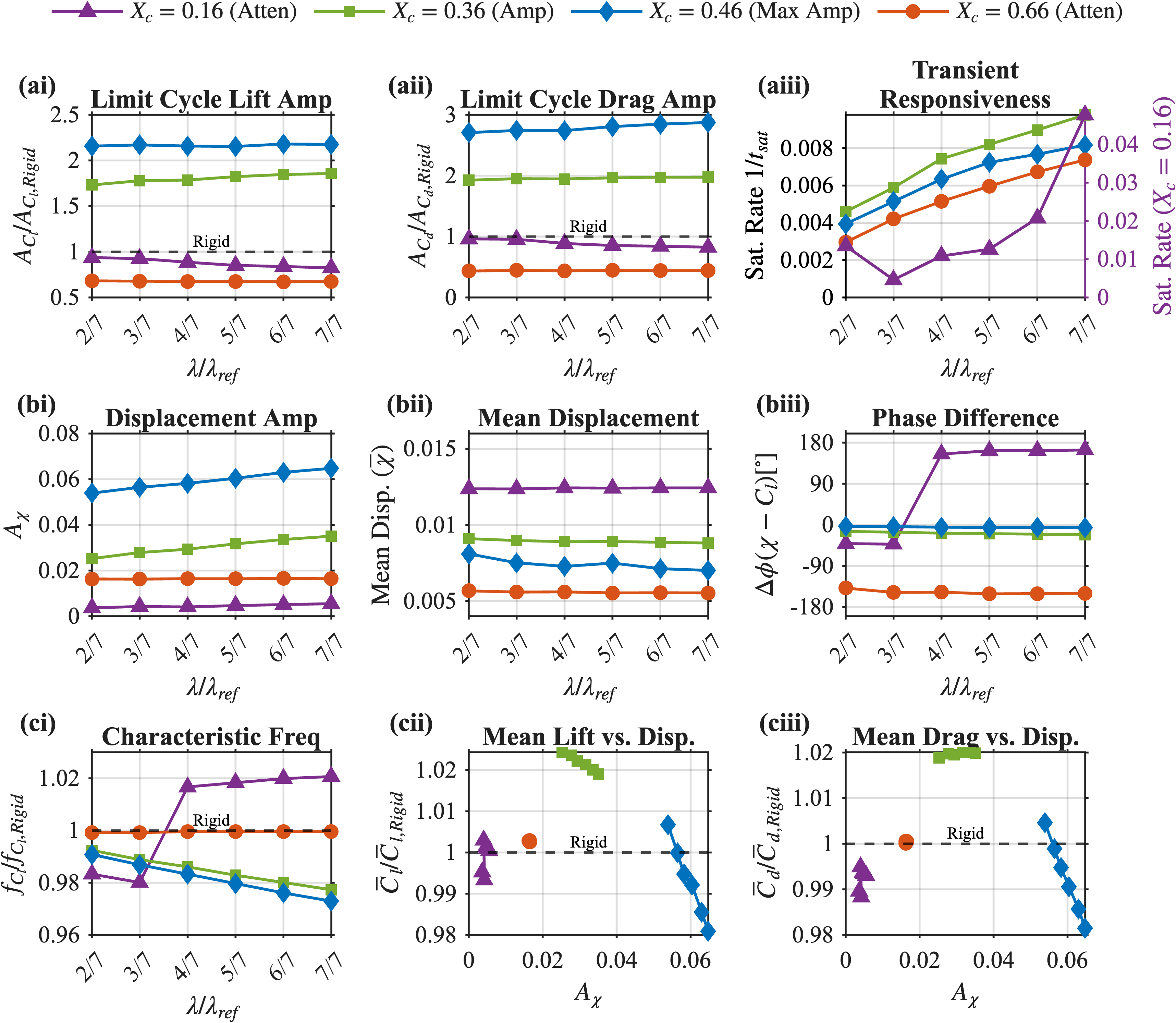}
    \caption{Effect of $\lambda/\lambda_{\mathrm{ref}}$, at the four representative chordwise positions $X_c = 0.16, 0.36, 0.46, 0.66$, on the following variables: \textbf{(ai)} normalized lift amplitude, (\textbf{(aii)}) normalized drag amplitudes, \textbf{(aiii)} the transient saturation rate, $1/t_{\mathrm{sat}}$ (note the different y-axis on the right for $X_c=0.16$, to visualize all data on the same plot),
    \textbf{(bi)})the displacement oscillation $A_\chi$, 
    \textbf{(bii)} the mean displacement $\bar{\chi}$, 
    \textbf{(biii)} the phase difference between the displacement and lift, 
    \textbf{(ci)} the normalized lift oscillation frequency, 
    \textbf{(cii)} mean lift as a function of displacement oscillation amplitude $A_\chi$, and \textbf{(ciii)} mean drag as a function of  displacement oscillation amplitude $A_\chi$.}
    \label{fig:4_1a}
\end{figure}

Figure~\ref{fig:4_1a} presents the dependence of the various characteristics of the limit cycle vortex shedding on $\lambda$ at all four representative PSub chordwise positions. For all four positions, the limit cycle amplitudes of lift and drag (Figure~\ref{fig:4_1a} (ai) and (aii)) remain relatively constant across all $\lambda$, confirming that the limit cycle amplitude is largely dictated by the PSub chordwise position. However, the transient saturation rate $1/t_{\mathrm{sat}}$ (Figure~\ref{fig:4_1a} (aiii)) scales monotonically with $\lambda$ for $X_c = 0.36, 0.46$, and $0.66$, with a notably similar trend regardless of whether the PSub is in an amplification or attenuation regime. In contrast, the transient responsiveness at $X_c = 0.16$ (Figure~\ref{fig:4_1a} (aiii), right axis) shows a non-monotonic feature, with $\lambda/\lambda_{\mathrm{ref}} = 2/7$ exceeding $3/7$ before increasing through higher $\lambda$ values. Overall, the results suggest that $\lambda$ has significant impact on the \emph{rate} that the PM--FSI reaches the limit cycle, and relatively little influence on limit cycle lift and drag oscillation amplitudes.


The oscillation amplitude of the displacement in the limit cycle $A_\chi$ (Figure~\ref{fig:4_1a} (bi)) scales approximately linearly with $\lambda$ for the amplification cases ($X_c = 0.36, 0.46$), while the attenuation cases ($X_c = 0.16, 0.66$) show nearly constant displacement oscillation across the $\lambda$. Similarly, the mean displacement $\bar{\chi}$ (Figure~\ref{fig:4_1a} (bii)) depends on $\lambda$ for the amplification cases but is independent of $\lambda$ for the attenuation cases. It is worth noting that at $X_c = 0.46$ higher $\lambda$ produces both larger displacement oscillation amplitude and smaller mean displacement, meaning the compliant surface intrudes further toward the pressure side and stays there for a longer fraction of the cycle. This difference in how the PSub affects the pressure side correlates with the larger mean lift dependence on displacement at $X_c = 0.46$.

The phase difference between the PSub displacement and lift, $\Delta\phi(\chi, C_l)$, (Figure~\ref{fig:4_1a} (biii)) is nearly invariant to $\lambda$ for $X_c = 0.36, 0.46$, and $0.66$. In contrast, $X_c = 0.16$ undergoes a discrete shift from $\approx -45^\circ$ in the low-$\lambda$ regime to $\approx 170^\circ$ (nearly out-of-phase) in the high-$\lambda$ regime. The characteristic frequency of the lift signal (Figure~\ref{fig:4_1a} (ci)) decreases with $\lambda$ for the amplification cases at $X_c = 0.36$ and $0.46$, is nearly constant for $X_c = 0.66$, and exhibits a binary jump from $\sim 2\%$ below rigid to $\sim 2\%$ above rigid across the threshold at $X_c = 0.16$. Despite the binary character of the $X_c = 0.16$ response, the high-$\lambda$ regime aligns with the trend observed in Section \ref{subsec:3.2}: in the attenuation cases ($X_c = 0.16$ in the high-$\lambda$ regime and $X_c = 0.66$ across all $\lambda$), PSub displacement and lift are nearly out-of-phase and the lift frequencies remain close to or above the rigid baseline, while in the amplification cases ($X_c = 0.36, 0.46$) PSub displacement and lift are in-phase and the lift frequencies are lower than the rigid case. These results suggest that larger displacement amplitudes in amplification cases at higher $\lambda$ act as an added mass to the coupled dynamical system, as discussed in Section \ref{subsec:3.2}, which reduces the characteristic frequency. Conversely, the out-of-phase coupling could be associated with an effective added stiffness \citep{leonard2001aspects,shiels2001flow} or effective negative added mass \citep{williamson2004vortex} from the surrounding flow, contributing to the maintained or slightly higher frequency relative to the rigid case. Further studies could confirm these hypotheses.

The mean lift and drag across the six $\lambda$ values for the four representative chordwise positions (Figure~\ref{fig:4_1a} (cii) and (ciii)) are plotted against $A_\chi$ rather than $\lambda$, as $\lambda$ appears to influence the mean lift and drag primarily through $A_\chi$, rather than directly. Plotting against $A_\chi$ therefore illustrates the impact on the mean lift and drag more directly. At $X_c = 0.46$, mean lift and drag decrease with increasing $A_\chi$. As mentioned earlier, at higher $\lambda$, the compliant surface intrudes further toward the pressure side and stays there for a longer fraction of the cycle. This prolonged disruption of the pressure side flow correlates with greater reduction of mean lift and drag. At $X_c = 0.36$, the mean lift and drag dependence on $A_\chi$ is weaker than at $X_c = 0.46$, consistent with the weaker dependence of $A_\chi$ on $\lambda$. The mean lift shows a slight decrease with increasing $A_\chi$ and the mean drag shows nearly constant, non-monotonic dependence on $A_\chi$. At $X_c = 0.66$, both the mean lift and drag collapse to a single point across $\lambda$ variation, consistent with $A_\chi$ and $\bar{\chi}$ being independent of $\lambda$.  $X_c = 0.16$ shows distinct behavior compared to other chordwise positions: even with nearly identical $A_\chi$ across $\lambda$, the mean lift cluster around two distinct points depending on $\lambda$: slightly lower mean lift and drag for the low-$\lambda$ and higher for the high-$\lambda$. This is notable because, for this chordwise position, $A_\chi$ alone does not explain the shift in mean loads at this position, suggesting that the mean lift and drag at this position may be governed by the phase difference and/or the characteristic frequency of the coupled FSI. This behavior is not fully resolved in the present study and merits further study.

These results across the four chordwise positions at different $\lambda$ values show that $\lambda$ primarily governs the transient responsiveness of the PM--FSI, with comparatively little impact on the limit cycle amplitudes themselves. However, examining the various fluid and structural variables reveals that this $\lambda$ dependence is dictated by chordwise position, separating into two distinct behaviors: the three positions close to the near-surface $u'_{x,rms}$ peak ($X_c = 0.36, 0.46, 0.66$) show a smooth, largely monotonic dependence on $\lambda$, while the separation-onset position ($X_c=0.16$ shows a qualitatively different, threshold-like dependence between low-$\lambda$ and high-$\lambda$ regimes. This distinguishes $X_c = 0.16$ from the other three positions, which we analyze further below.

\subsection{Flow Field Analysis and Discussion}
\label{subsec:4.2}

This subsection examines both the unsteady (RMS) and mean flow fields to investigate how $\lambda$ modulates the PM--FSI. The analysis focuses on the streamwise velocity $u_x$, and note corresponding observations for $\omega_z$ and $u_y$ are consistent with the conclusions drawn here.

\begin{figure}[H]
    \centering
    \includegraphics[width=0.9\textwidth]{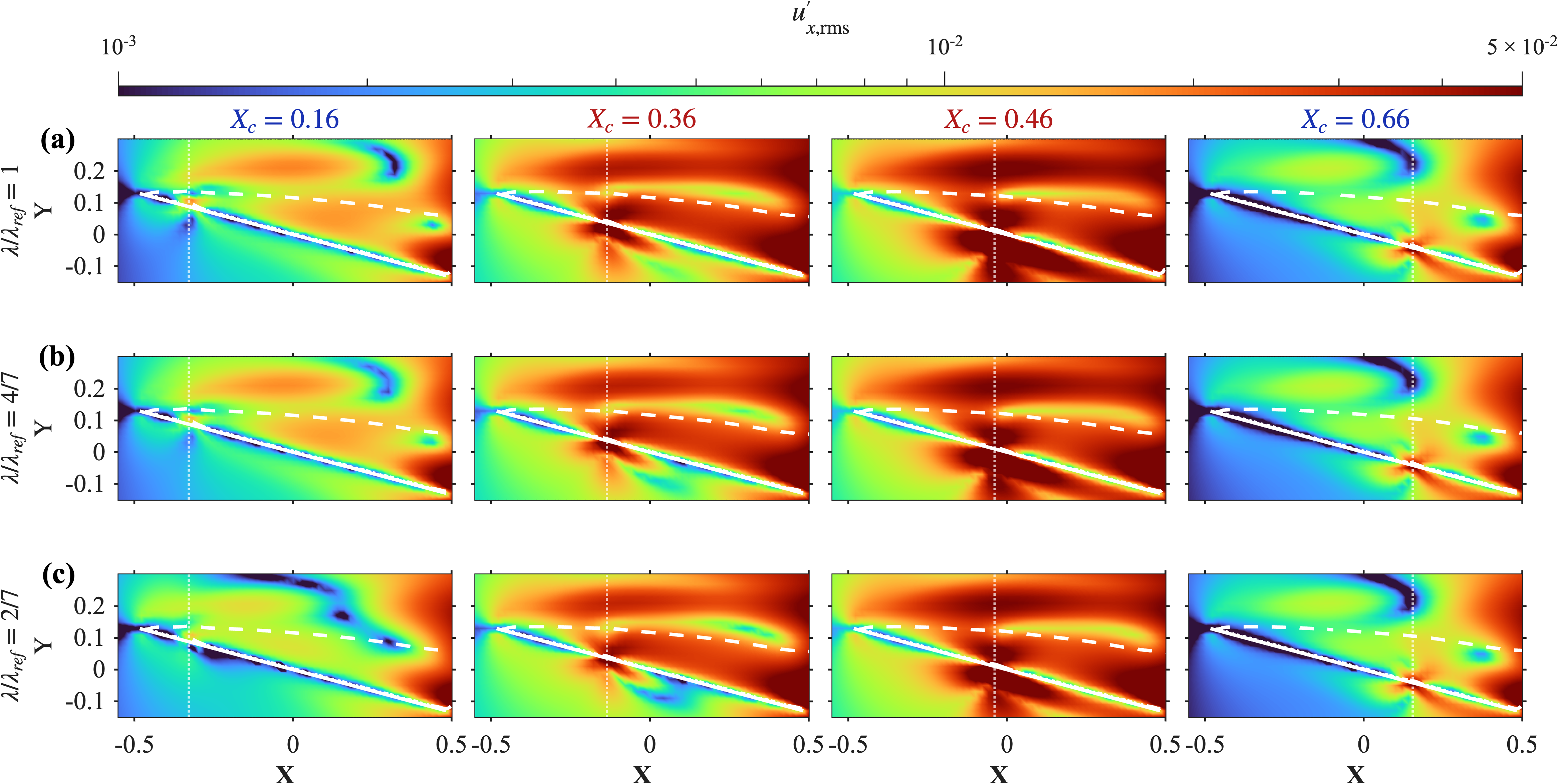}
    \caption{RMS streamwise velocity for 4 Psub positions at (a) $\lambda/\lambda_{\mathrm{ref}} = 1$, (b) $\lambda/\lambda_{\mathrm{ref}} = 4/7$, (c) $\lambda/\lambda_{\mathrm{ref}} = 2/7$. Dashed white lines indicate the $\bar{u}_x = 0$ boundary for each case. Dotted white lines indicate the X location of the PSub.}
    \label{fig:4_2_ux_RMS}
\end{figure}

\begin{figure}[H]
    \centering
    \includegraphics[width=0.9\textwidth]{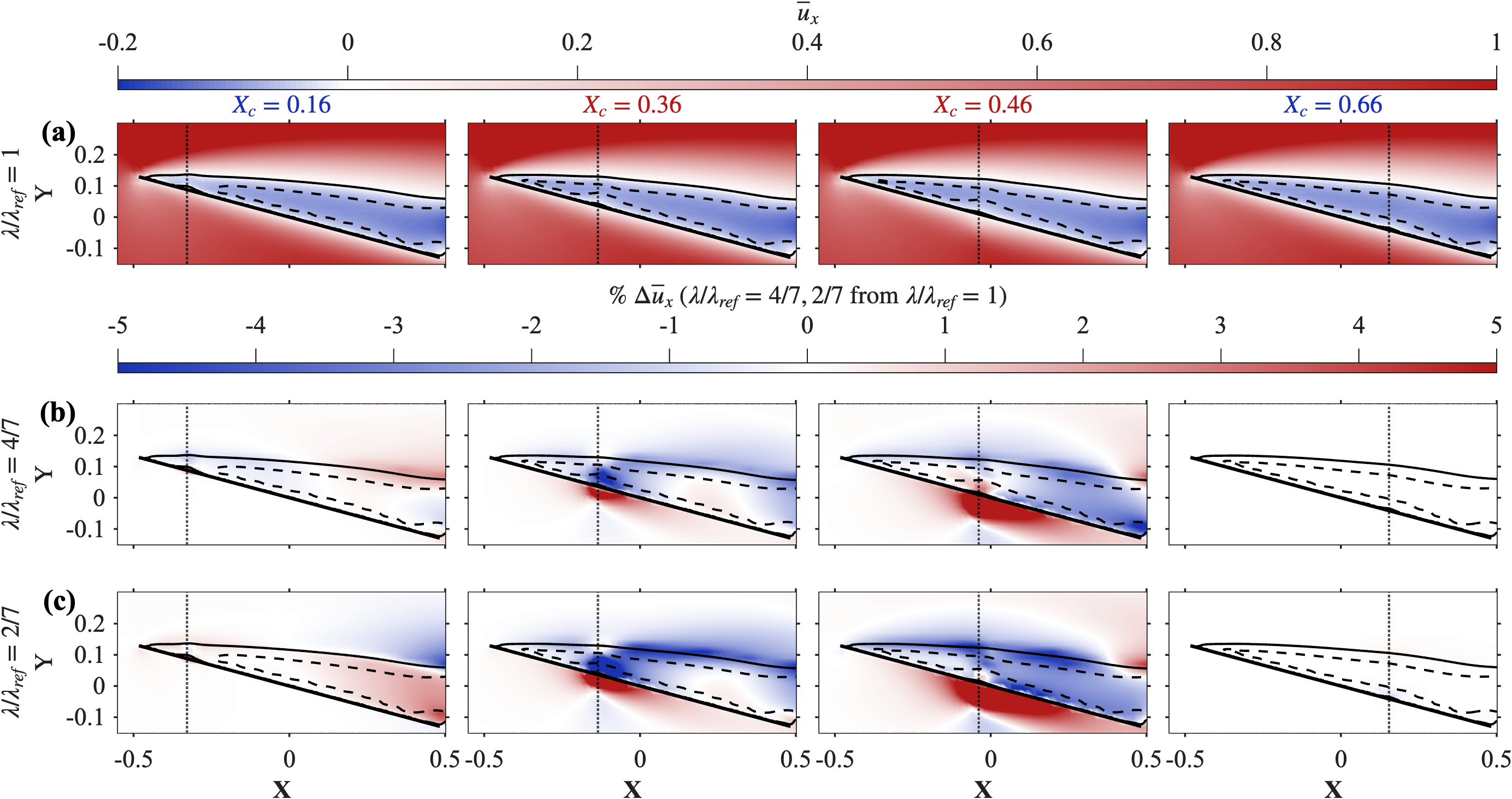}
    \caption{(a) Mean streamwise velocity for the high $\lambda$ case ($\lambda/\lambda_{\mathrm{ref}} = 1$) for 4 PSub positions. Solid black lines indicate $\bar{u}_x = 0$, and dashed black lines indicate $\bar{u}_x = -0.08$. (b-c) Percentage difference in mean streamwise velocity $\%\Delta\bar{u}_x$ relative to the $\lambda/\lambda_{\mathrm{ref}} = 1$ reference for the four representative positions ($X_c = 0.16, 0.36, 0.46, 0.66$, columns) for (b) $\lambda/\lambda_{\mathrm{ref}} = 4/7$ and (c) $\lambda/\lambda_{\mathrm{ref}} = 2/7$ (rows).}
    \label{fig:4_2_ux_mean}
\end{figure}

Figure~\ref{fig:4_2_ux_RMS} shows $u'_{x,\mathrm{rms}}$ for the four representative positions at $\lambda/\lambda_{\mathrm{ref}} = 1, 4/7$, and $2/7$. For $X_c = 0.36, 0.46$, and $0.66$, the spatial pattern and magnitude of $u'_{x,\mathrm{rms}}$ remain similar across all three $\lambda$ values. In contrast, the unsteady flow field in $X_c = 0.16$ shows a distinct dependence on $\lambda$. The spatial signature of of $u'_{x,\mathrm{rms}}$ at $\lambda/\lambda_{\mathrm{ref}} = 4/7$ remains similar to $\lambda/\lambda_{\mathrm{ref}} = 1$, but at $\lambda/\lambda_{\mathrm{ref}} = 2/7$, the spatial signature changes noticeably, in addition to a change in magnitude. This distinct spatial change at low $\lambda$ suggests that the coupled PM--FSI at $X_c = 0.16$ is fundamentally different at this $\lambda$ value compared to the other positions, consistent with the threshold-like binary response observed in the limit cycle behavior at $X_c = 0.16$ (Fig.~\ref{fig:4_1a}).

Figure~\ref{fig:4_2_ux_mean} shows the mean $\bar{u}_x$ for $\lambda/\lambda_{\mathrm{ref}} = 1$ and the percentage difference in $\lambda/\lambda_{\mathrm{ref}} = 4/7$ and $2/7$ from $\lambda/\lambda_{\mathrm{ref}} = 1$. For $X_c = 0.36$, and $0.46$, similar regions in the flow field show increased or decreased $\bar{u}_x$. Only the magnitude of the change differs, larger change with decreasing $\lambda$. The recirculation contours ($\bar{u}_x = 0$ and $\bar{u}_x = -0.08$) still remain nearly identical across $\lambda$ at these positions as the percentage change in $\bar{u}_x$ is mostly below 5\% in the recirculation. These observations indicate that at $X_c = 0.36$, and $0.46$, $\lambda$ scales the magnitude of the mean flow change, though marginal, without changing where the flow is altered. $X_c = 0.46$ case shows significant pressure-side $\bar{u}_x$ change with $\lambda$, which correlates with the stronger dependence of mean lift on $\lambda$ at this position (Fig.~\ref{fig:4_1a}). At $X_c = 0.66$, the $u_x$ deviation is minimal across all $\lambda$ values, showing almost no dependence on $\lambda$. This is consistent with the negligible mean flow modification noted in Section \ref{subsec:3.3}.

In contrast, at $X_c = 0.16$ the spatial pattern of the mean flow change differs between $\lambda/\lambda_{\mathrm{ref}} = 4/7$ and $2/7$. The regions of increased and decreased $\bar{u}_x$ shift with $\lambda$, even though the $\bar{u}_x = 0$ and $\bar{u}_x = -0.08$ contours remain relatively constant. At this position, $\lambda$ changes not only the magnitude of the mean flow change but also its spatial signature. This also correlates with the threshold-like binary dependence on $\lambda$ observed in the limit cycle behavior at $X_c = 0.16$ (Fig.~\ref{fig:4_1a}).

This difference in how $\lambda$ affects the flow field at $X_c = 0.16$ compared to the other PSub positions may explain why the limit cycle at $X_c = 0.16$ shows a threshold-like binary dependence on $\lambda$, while those at the other positions vary either monotonically with $\lambda$ ($X_c = 0.36, 0.46$) or remain nearly constant ($X_c = 0.66$). When the PSub interacts with the unsteady recirculation already present in the rigid baseline (PSub positions such as $X_c = 0.36, 0.46, 0.66$), $\lambda$ primarily dictates the transient saturation rate, and the limit cycle characteristics either scale smoothly with $\lambda$ or are set by the PSub position. When the PSub is positioned at the onset of recirculation, as in the case of $X_c = 0.16$, $\lambda$ additionally influences how the PSub alters the recirculation, and thus the limit cycle. 

\section{Conclusions}
\label{sec:conclusions}

This work used strongly coupled high-fidelity simulations to compute the nonlinear FSI dynamics of a fully separated unsteady flow past a flat plate interacting with a phononic subsurface. Simulations were performed varying the position of the PSub along the plate ($X_c$) and the displacement amplitude envelope ($\lambda$). The results show that $X_c$ primarily determines whether the limit cycle lift and drag oscillations are amplified or attenuated through the PSub's spatial relationship to the surrounding flow instability. 
Specifically, four distinct regimes are identified as a function of the PSub chordwise position:
\begin{enumerate}
    \item \textit{Upstream attenuation} ($X_c = 0.16$): the PSub alters the flow at the onset of the recirculation, reducing lift and drag oscillation amplitude with a slight mean drag reduction.
    \item \textit{Upstream amplification} ($X_c = 0.36$): the PSub seeds perturbations into the recirculation zone without significantly affecting the pressure side flow, amplifying lift and drag oscillations while slightly increasing mean lift.
    \item \textit{Downstream amplification} ($X_c = 0.46$): the PSub amplifies lift and drag oscillation amplitude and additionally disrupts the pressure side, leading to a slight reduction in mean lift.
    \item \textit{Downstream attenuation} ($X_c = 0.66$): the PSub reduces lift oscillation amplitude by over 30\% and drag oscillation amplitude by over 50\% with less than 0.5\% change in mean lift and drag.
\end{enumerate}
 Results suggest that the amplification or attenuation induced by the PSub depends on its position relative to distinct spatial features of the underlying flow behaviors.  Specifically, the lift and drag oscillation amplitudes change sharply when the PSub is close to the near-surface $u'_{x,\mathrm{rms}}$ peak of the rigid baseline ($X_{c,\mathrm{peak}}=0.58$). Meanwhile, the upstream attenuation positions ($X_c=0.06, 0.16$), which are far from this peak but close to the onset of the recirculation, appear to alter the vortex shedding through the recirculation onset rather than through direct interaction with the peak near-surface near-surface $u'_{x,rms}$ peak $X_{c,\mathrm{peak}}$. 

The displacement amplitude envelope $\lambda$ primarily affects the transient responsiveness toward the limit cycle, dictating the time scale to reach the limit cycle behavior. In addition to the transient responsiveness, the limit cycle characteristics show slight dependence on $\lambda$ in a position-dependent manner: the dependence on $\lambda$ is monotonic in the amplification regimes ($X_c=0.36, 0.46$), remains nearly invariant in the downstream attenuation regime ($X_c=0.66$), and undergoes a threshold-like binary shift in the upstream attenuation regime ($X_c=0.16$).

The downstream attenuation case at $X_c = 0.66$ is particularly notable in terms of passive flow control: it causes significant reductions in both lift and drag oscillation amplitude, with minimal change in mean lift and drag. This demonstrates that a PSub can passively suppress unsteady aerodynamic loads through its chordwise placement with respect to the flow instabilities, without compromising the mean aerodynamic performance.

While the specific regimes and corresponding $X_c$ values identified here are tied to the specific flow condition studied, the results here reveal a more general principle of PM--FSI with vortex shedding: the flow is sensitive to the position of a PSub relative to the spatial structure of the underlying flow behavior. Further, different PSub positions induce different changes to the response of the coupled dynamical system.  These results also reveal the complex nature of FSI between phononic materials and fluid flows, that can only be studied when the PSub is strongly coupled to the nonlinear flow.

\section*{Acknowledgments} 
This material is based upon work supported by the Air Force Office of Scientific Research under award number FA955021-1-0182 and award number FA9550-23-1-0299.

\appendix
\section{Mass and Stiffness Matrices of the Diatomic chain PM }\label{app:mk_matrix}

The mass and stiffness matrices of the $N$ unit cells of the grounded diatomic chain PM shown in Fig.~\ref{fig:problem_config}(b) are given as:

\begin{equation*}
    \mathbf{M}=\begin{bmatrix}
                    m_1 & 0 & 0 & \cdots & 0\\
                    0 & m_2 & 0 & \cdots & 0\\
                    0 & 0 & m_1 & \ddots & \vdots\\
                    \vdots & \vdots & \ddots & \ddots & 0\\
                    0 & 0 & \cdots & 0 & m_2
                \end{bmatrix}_{2N\times2N},
\end{equation*}
\begin{equation*}
    \mathbf{K}=\begin{bmatrix}
                    k & -k & 0  & \dots & 0\\
                    -k & 2k+k_\mathrm{g} & -k & \dots & 0\\
                    0 & -k  & 2k & \ddots & \vdots\\
                    \vdots & \vdots & \vdots & \ddots  & -k\\
                    0 & 0 & \dots & -k & k+k_\mathrm{g}
                \end{bmatrix}_{2N\times2N}.
\label{eq:mk_matrix}
\end{equation*}

\section{Grid Convergence Study}
\label{app:grid_convergence}

A grid convergence study was conducted to verify the spatial and temporal resolution of the computational domain. Table~\ref{tab:grid_convergence} summarizes the robust mean lift coefficient ($C_l$) and the characteristic vortex shedding frequency ($f_{C_l}$) across seven grid refinement levels. Grid 1 represents the finest resolution and serves as the baseline for the percentage difference calculations. The results demonstrate that the force and frequency metrics for the selected working grid (Grid 2) are well converged, with deviations of less than 1\% from the finest grid.

\begin{table}[H]
    \centering
    \caption{Grid convergence summary. The percentage differences are calculated with respect to the finest grid (Grid 1).}
    \label{tab:grid_convergence}
    \begin{tabular}{l c c c c c c c}
        \toprule
        \textbf{Grid} & \textbf{$\Delta x$} & \textbf{$\Delta t$} & \textbf{Total domain} & \textbf{$\bar{C}_l$} & \textbf{$\%|\Delta\bar{C}_l|$} & \textbf{$f_{C_l}$} & \textbf{$\%|\Delta f_{C_l}|$} \\
        \midrule
        Grid 7 & 0.0400 & 0.0080 & $[-5.5, 6.5] \times [-4, 4]$     & 0.8508 & 3.441 & 0.4854 & 19.53 \\
        Grid 6 & 0.0200 & 0.0040 & $[-5.5, 6.5] \times [-4, 4]$     & 0.8571 & 4.215 & 0.5550 & 7.984  \\
        Grid 5 & 0.0100 & 0.0020 & $[-5.5, 6.5] \times [-4, 4]$     & 0.8371 & 1.782 & 0.5879 & 2.530  \\
        Grid 4 & 0.0050 & 0.0010 & $[-23.5, 24.5] \times [-16, 16]$ & 0.8035 & 2.300 & 0.5898 & 2.213  \\
        Grid 3 & 0.0050 & 0.0010 & $[-11.5, 12.5] \times [-8, 8]$   & 0.8107 & 1.4334 & 0.5927 & 1.739  \\
        Grid 2 & 0.0050 & 0.0010 & $[-5.5, 6.5] \times [-4, 4]$     & 0.8273 & 0.5860 & 0.5989 & 0.7115  \\
        Grid 1 & 0.0025 & 0.0005 & $[-5.5, 6.5] \times [-4, 4]$     & 0.8225 & --     & 0.6032 & --      \\
        \bottomrule
    \end{tabular}
\end{table}

\newpage
\bibliographystyle{elsarticle-num-names}
\bibliography{references}

\end{document}